\documentclass[journal,12pt,draftclsnofoot,onecolumn]{IEEEtran}

\usepackage{cite}

\usepackage{makeidx,epsfig}
\usepackage{setspace,graphicx,multirow}

\usepackage{amsfonts,latexsym,amssymb,amsthm}

\usepackage{graphicx}
\usepackage{epstopdf}

\usepackage{caption3} 
\DeclareCaptionOption{parskip}[]{} 
\usepackage[small]{caption}

\usepackage{subcaption}

\usepackage{array}

\usepackage[cmex10]{amsmath}

\usepackage{url}

\usepackage{multirow}
\usepackage{hhline}

\usepackage{amsmath}
\usepackage{algorithm}
\usepackage[]{algpseudocode}
\usepackage{varwidth}

\makeatletter
\def\BState{\State\hskip-\ALG@thistlm}
\makeatother

\usepackage{algcompatible}

\usepackage{fixltx2e}

\newcommand*\diff{\mathop{}\!\mathrm{d}}

\usepackage{enumitem}

\usepackage{color}

\begin{document}
\bibliographystyle{IEEEtran}
\bstctlcite{IEEEexample:BSTcontrol}

\title{UAV-Aided Offloading for Cellular Hotspot}

\author{Jiangbin~Lyu,~\textit{Member,~IEEE},
        Yong~Zeng,~\textit{Member,~IEEE},
        and~Rui~Zhang,~\textit{Fellow,~IEEE}%
\thanks{Part of this work has been presented at IEEE Globecom 2017, Singapore \cite{UAVGBSglobecom}.}
\thanks{J. Lyu is with School of Information Science and Engineering, Xiamen University, China 361005 (e-mail: ljb@xmu.edu.cn).}
\thanks{Y. Zeng and R. Zhang are with the Department of Electrical and Computer Engineering, National University of Singapore, Singapore 117583 (email: \{elezeng, elezhang\}@nus.edu.sg).}%
}

\maketitle

\begin{abstract}
In conventional terrestrial cellular networks, mobile terminals (MTs) at the cell edge often pose a performance bottleneck due to their long distances from the serving ground base station (GBS), especially in hotspot period when the GBS is heavily loaded. This paper proposes a new hybrid network architecture by leveraging the use of unmanned aerial vehicle (UAV) as an aerial mobile base station, which flies cyclically along the cell edge to offload data traffic for cell-edge MTs. We aim to maximize the minimum throughput of all MTs by jointly optimizing the UAV's trajectory, bandwidth allocation and user partitioning. We first consider orthogonal spectrum sharing between the UAV and GBS, and then extend to spectrum reuse where the total bandwidth is shared by both the GBS and UAV with their mutual interference effectively avoided. Numerical results show that the proposed hybrid network with optimized spectrum sharing and cyclical multiple access design significantly improves the spatial throughput over the conventional GBS-only network; while the spectrum reuse scheme provides further throughput gains at the cost of slightly higher complexity for interference control. Moreover, compared to the conventional small-cell offloading scheme, the proposed UAV offloading scheme is shown to outperform in terms of throughput, besides saving the infrastructure cost.
\end{abstract}

\section{Introduction}


%
%
%
%
%
%

With their high mobility and ever-reducing cost, unmanned aerial vehicles (UAVs) are expected to play an important role in future wireless communication systems. 
There are assorted appealing applications by leveraging UAVs for wireless communications \cite{ZengUAVmag}, such as UAV-enabled ubiquitous coverage or drone small cells (DSCs) \cite{CyclicalLyu,WuQingQingUAVTWC,PlacementLyu,3Dplacement,Halim3Dnumber,HalimRAN,UAVd2d,PlacementCirclePacking,LAPlosProbability,ZhangWeiSmallCellJSAC}, UAV-enabled mobile relaying \cite{UAVrelay, ZengMobileRelay, UAVwifiExperiment} and UAV-enabled information dissemination/data collection \cite{UAVDataCollection,YongCompletionTime,ZhanChengCollection}. In particular, for UAV-enabled ubiquitous coverage, the UAV is deployed to assist the existing terrestrial communication system in providing seamless wireless coverage. Two typical use scenarios are rapid service recovery after ground infrastructure malfunction \cite{UAVpublicSafety} and cellular traffic offloading from overloaded ground base stations (GBSs) in, e.g., hotspot areas. Note that the latter case has been identified as one of the five key scenarios that need to be effectively addressed by the fifth-generation (5G) wireless systems \cite{Hotspot5G}.

The offloading issue for cellular hotspot can be partly addressed via existing technologies such as WiFi offloading \cite{WiFiOffloading} or small cell \cite{SmallCell5G}, among others. 
However, these solutions usually require deploying new fixed access points/GBSs, which could be cost-ineffective for scenarios with highly dynamic and diversified traffic demand such as open air festivals and other public events with temporarily high user density.
In such scenarios, UAV-aided cellular offloading provides a promising alternative solution to address the cellular hotspot issue, of which the main cost such as the energy and aircraft cost can be lower than building new ground infrastructure. Furthermore,
UAV-aided cellular offloading offers promising advantages compared to the conventional cellular network with fixed GBSs,
such as the ability for on-demand and swift deployment, more flexibility for network reconfiguration, and better communication channels between the UAV and ground mobile terminals (MTs) due to the dominant line-of-sight (LoS) links.
Moreover, the UAV mobility provides additional design degrees of freedom via trajectory optimization \cite{ZengEnergyEffcient}.



\begin{figure}
\centering
   \includegraphics[width=0.8\linewidth]{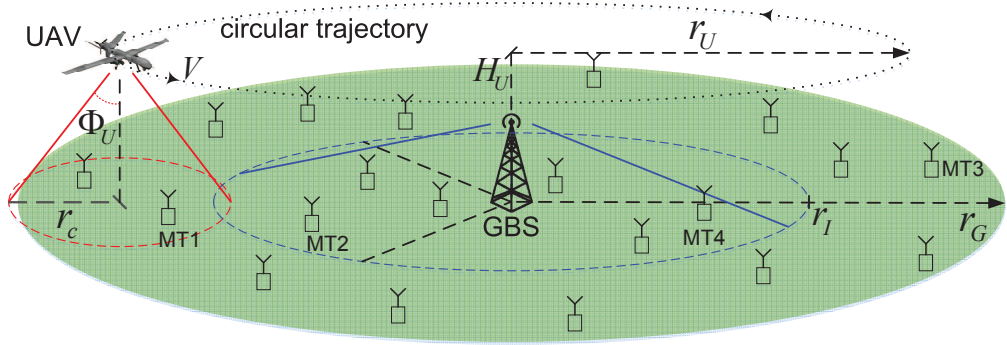}
\caption{UAV-aided cellular offloading.\vspace{-2ex}}\label{Schematic}
\end{figure}

In traditional terrestrial cellular networks, the cell-edge MTs often suffer from poor channel conditions due to their long distances from their associated GBS.
As a result, with a limited total bandwidth available for each cell, these cell-edge MTs would require either more bandwidth and/or higher transmit power in order to achieve the same performance as other non-cell-edge MTs, which thus pose a fundamental performance bottleneck for the cellular system, especially for hotspot period when the GBS is heavily loaded.
To tackle this issue, we propose in this paper a new hybrid cellular network architecture based on the technique of UAV-aided cellular offloading. The proposed hybrid network architecture consists of a conventional GBS and an additional UAV serving as an aerial mobile BS to jointly serve the MTs in each cell. As shown in Fig. \ref{Schematic}, the UAV flies cyclically along the cell edge to serve the cell-edge MTs and thereby help offloading the traffic from the GBS.
Accordingly, the MTs in the cell are partitioned into cell-edge and non-cell-edge MTs, which are served by the UAV and GBS, respectively.
We assume that the UAV flies at a fixed altitude following a circular trajectory with a certain radius centered at the GBS, and communicates with its associated cell-edge MTs in a cyclical time-division manner \cite{CyclicalLyu}. Specifically, at any time instant, only those cell-edge MTs that are sufficiently close to the UAV are scheduled to communicate with the UAV.
Compared to the small cell technology where usually a large number of small cells need to be deployed in different fixed locations in the cell,
the UAV-enabled cyclical multiple access scheme essentially shortens the communication distance with all cell-edge users by exploiting the UAV's mobility, and hence it is anticipated to significantly reduce the deployment cost and improve the system throughput.




With the proposed hybrid network architecture applied to a single-cell system, we study the problem of maximizing the minimum (common) throughput of all MTs in the cell, so that each MT achieves a fair common throughput.
Specifically, the main contributions of this paper are summarized as follows.

\begin{itemize}
\item First,
we consider the case of orthogonal spectrum sharing between the GBS and UAV, where the total available bandwidth is partitioned into two orthogonal parts to be allocated to the UAV and GBS, respectively. 
Three key parameters are then jointly designed, namely the bandwidth allocation and the user partitioning between the UAV and GBS, as well as the UAV's circular trajectory radius.
The joint optimization problem is non-convex and challenging to be directly solved. To tackle this problem, we first optimize the UAV's trajectory radius for given bandwidth allocation and user partitioning. Then we jointly optimize the bandwidth allocation and user partitioning to maximize the common throughput of all MTs.

\item Second, 
we extend our analysis to the spectrum reuse case where the whole spectrum pool is shared by both the GBS and UAV for concurrent communications. In this case, their mutual interference is a key issue and we propose effective methods to suppress  the interference by leveraging the use of directional antennas at the UAV and adaptive directional transmission at the GBS.
Compared to the orthogonal spectrum sharing scheme, the spectrum reuse scheme further improves the spectrum efficiency and thus the common throughput, at the cost of more complexity in practical implementation for the interference avoidance between the UAV and GBS transmissions.

\item Finally, extensive numerical results are provided to validate our analytical results. It is found that the proposed hybrid network with optimized design greatly improves the spatial throughput over the traditional network with the GBS only. As a result, the proposed UAV-aided cellular offloading scheme can support higher user density under the same target rate requirements for each user, which thus provides a promising solution to address the cellular hotspot issue.
Furthermore, it is shown that the joint optimization of spectrum sharing, multiple access, and UAV trajectory design is essential to achieve the optimum throughput of the proposed UAV-assisted hybrid network, for both cases with orthogonal spectrum sharing and non-orthogonal spectrum reuse. 
Moreover, the proposed scheme is also compared with the conventional cell-edge throughput enhancement scheme by deploying a number of micro/small cells to help offload data traffic for cell-edge users. The simulation results show that the proposed UAV offloading scheme with only one single UAV/mobile BS significantly outperforms the micro-cell offloading scheme in terms of throughput, besides saving the infrastructure cost.
\end{itemize}

The rest of this paper is organized as follows.
The system model and the proposed UAV-enabled hybrid network architecture are given in Section \ref{SectionModel}.
The optimized designs for maximizing the minimum throughput with orthogonal spectrum sharing scheme and spectrum reuse scheme are presented in Section \ref{SectionFormulation} and Section \ref{SectionReuse}, respectively.
Section \ref{SectionReuse} also provides discussions on the relaxation of the modeling assumptions and some practical implementation issues.
Numerical results are provided in Section \ref{SectionSimulation}.
Finally, we conclude the paper in Section \ref{SectionConclusion}.

\section{System Model}\label{SectionModel}

As shown in Fig. \ref{Schematic}, we consider a single-cell wireless communication system with a GBS and a UAV jointly serving a group of MTs on the ground. In this paper, we consider the downlink communication from the GBS/UAV to the MTs, whereas the obtained results can be similarly applied to the uplink communication as well. 
Assume that the MTs are uniformly and randomly distributed with a given density $\lambda$ in the cell of cell radius $r_G$ and centered at the GBS location; thus, the total number of MTs on average is $K=\pi r_G^2\lambda$. Denote the set of MTs as $\mathcal{K}=\{1,2,\cdots,K\}$.
The MTs are partitioned into two disjoint groups, $\mathcal{K}_G$ and $\mathcal{K}_U$, based on a distance threshold $r_I$ to the GBS, where $\mathcal{K}_G$ denotes the set of MTs in the \textit{inner disk} region of radius $r_I$, and $\mathcal{K}_U$ denotes the remaining MTs in the exterior \textit{ring} region.
We assume that the MTs in $\mathcal{K}_G$ (e.g., MTs 2 and 4 in Fig. \ref{Schematic}) are associated with the GBS for communications, while those in $\mathcal{K}_U$ (e.g., MTs 1 and 3) are served by the UAV via the cyclical multiple access scheme \cite{CyclicalLyu}.
Hence, there are on average $K_G\triangleq |\mathcal{K}_G|=\pi\lambda r_I^2$ MTs associated with the GBS, and $K_U\triangleq |\mathcal{K}_U|=\pi \lambda (r_G^2-r_I^2)$ MTs to be served by the UAV, where $|\cdot|$ denotes the cardinality of a set.
For simplicity, we assume that an ideal wireless backhaul between the UAV and GBS exists, which operates in a separate band. Several technologies such as millimeter wave and free space optical communications can be good candidates for realizing high-speed wireless backhaul between the UAV and GBS, thanks to the favorable communication channel with strong LoS link.


We assume that the UAV flies at a fixed altitude $H_U$, which could correspond to the minimum value required for safety considerations such as terrain or building avoidance.
We also assume that the UAV flies at a constant speed $V$ following a circular trajectory whose projection on the ground is centered at the GBS. Denote the radius of the UAV trajectory as $r_{U}$ and its period as $T$, i.e., the UAV position repeats every $T$ seconds, as shown in Fig.~\ref{Schematic}.
Then we have $T=2\pi r_U/V$.
Note that the circular trajectory is considered since it not only enables the UAV to serve the cell-edge users in a periodic manner, but is also energy-efficient for the UAV flying \cite{ZengEnergyEffcient}.
For any time instant $t$, let $\mathcal{K}_U(t)\subseteq\mathcal{K}_U$ denote the set of cell-edge MTs that are scheduled for communication with the UAV.
Since each MT has the best communication link when the UAV flies close to it, it is intuitive to schedule the nearest MTs from the current UAV position to communicate with the UAV, in order to maximize the system throughput.
Motivated by this, we propose a simple time-division based cyclical multiple access scheme, where different cell-edge MTs are scheduled to communicate with the UAV in a cyclical time-division manner to exploit the good channel when the UAV flies close to each of them.

Next, we discuss the channel models for UAV-MT and GBS-MT communications, respectively.
We assume that the UAV is equipped with a directional antenna, whose azimuth and elevation half-power beamwidths are both $2\Phi_U$ radians (rad) with $\Phi_U\in(0,\frac{\pi}{2})$. Furthermore, the corresponding antenna gain in direction $(\phi,\varphi)$ can be practically approximated as
\begin{align}\label{UAVantenna}
G_U(\phi,\varphi)=
\begin{cases}
G_0/\Phi_U^2, & \textrm{$-\Phi_U\leq \phi\leq \Phi_U$, $-\Phi_U\leq \varphi\leq \Phi_U$;}\\
g_0\approx 0, & \ \textrm{otherwise,}
\end{cases}
\end{align}%
where $G_0=\frac{30000}{2^2}\times(\frac{\pi}{180})^2\approx 2.2846$; $\phi$ and $\varphi$ denote the azimuth and elevation angles, respectively \cite{balanis2016antenna}\cite{HaiyunAntenna}. Note that in practice, $g_0$ satisfies $0<g_0 \ll G_0/\Phi_U^2$, and for simplicity we assume $g_0= 0$ in this paper.
On the other hand, we assume that each MT is equipped with an omnidirectional antenna of unit gain.
Thus, the disk region centered at the UAV's projection on the ground with radius $r_c=H_U\tan\Phi_U$ corresponds to the ground \textit{coverage area} by the antenna main lobe of the UAV, as shown in Fig. \ref{Schematic}. By properly adjusting the beamwidth $\Phi_U$, we assume that the coverage radius $r_c$ is appropriately set so that the scheduled MTs $\mathcal{K}_U(t)$ are guaranteed to lie within the coverage area of the UAV at time $t$.
On the other hand, an increase in $\Phi_U$ would reduce the antenna gain of the main lobe, as shown in \eqref{UAVantenna}.
Thus, the beamwidth $\Phi_U$ or equivalently the scheduled MTs $K_U(t)$ over time should be carefully designed.

We consider that the UAV-MT communication channels are dominated by LoS links. 
Though simplified, the LoS model offers a good approximation for practical UAV-MT channels, which is also one of the main motivations to utilize UAVs for wireless communication. Recent field experiments by Qualcomm \cite{QualCommDroneReport} have verified that the UAV-to-ground channel is indeed dominated by the LoS link for UAVs flying above a certain altitude.
We assume that the Doppler effect due to the UAV's mobility is perfectly compensated at all the MT receivers\footnote{In this paper, the UAV follows a simple circular trajectory with a fixed flying speed, thus the Doppler effect exhibits a certain cyclical pattern and hence can be more easily estimated and compensated.}.
Therefore, the channel power gain from the UAV to MT $k$ at time $t$ follows the free-space path loss model given by
\begin{equation}\label{PathLoss}
h_k(t)=\frac{\beta_0}{d_k^2(t)+H_U^2}, 0\leq t\leq T,
\end{equation}
where $\beta_0=(\frac{4\pi f_c}{c})^{-2}$ denotes the channel power gain at a reference distance of 1 meter (m), with $f_c$ denoting the carrier frequency and $c$ denoting the speed of light; and $d_k(t)$ is the horizontal distance between the UAV and MT $k$ at time $t$.

On the other hand, for GBS-MT communications,
we assume that the GBS has a fixed antenna gain for transmission, denoted by $G_G\geq 1$. In practice, the GBS could be equipped with an omnidirectional antenna, or multiple sectorized antennas with non-overlapping directional transmissions.
Furthermore, we assume a fading channel between the GBS and MTs, which consists of distance-dependent path-loss with path-loss exponent $n\geq 2$ and an additional random term accounting for small-scale fading.
Therefore, the channel power gain from the GBS to MT $k$ can be modelled as
$g_k=\bar g_k\zeta_k$, where $\bar g_k\triangleq\alpha_0 (H_G^2+r^2)^{-n/2}$ is the average channel power gain, with $\alpha_0=(\frac{4\pi f_c}{c})^{-2}$ denoting the average channel power gain at a reference distance of 1 m, $r$ denoting the horizontal distance between the GBS and MT $k$, and $H_G$ denoting the height of the GBS;
 and $\zeta_k\sim \textrm{Exp}(1)$ is an independent and identically distributed (i.i.d.) exponential random variable with unit mean accounting for the small-scale Rayleigh fading.

In this paper, we investigate two practical spectrum sharing models for the UAV and GBS, i.e., orthogonal spectrum sharing and non-orthogonal spectrum reuse.
In the orthogonal sharing case, the UAV and GBS are allocated with orthogonal spectrum respectively, and thus there is no interference between the UAV-MT and GBS-MT communications. 
By contrast, in the spectrum reuse case, the common spectrum pool is shared by both the GBS and UAV for concurrent transmissions, provided that their mutual interference is effectively suppressed. 
With directional/sectorized antennas, such interference can be avoided in practice by leveraging the joint use of directional antenna at the UAV and adaptive directional transmission at the GBS.
For example, in Fig. \ref{Schematic}, the GBS-MT4 and UAV-MT1 links can use the same frequency band at the same time without mutual interference if non-overlapping directional transmissions of the GBS and UAV are employed.
Note that spectrum reuse is a more general model than orthogonal sharing, which improves the spectrum efficiency but is also more complicated to design and implement in practice.

We assume that the total available bandwidth is $W$ Hz. In the orthogonal sharing case, denote the portion of bandwidth allocated to the UAV as $\rho$, with $0\leq\rho\leq 1$. 
Assume that the bandwidth allocated to the UAV is equally shared among the MTs associated with the UAV at each time, i.e., each MT $k\in\mathcal{K}_U(t)$ is allocated with an effective bandwidth of $b_U(t)W$, with $b_U(t)\triangleq\rho /|\mathcal{K}_U(t)|$ denoting the normalized bandwidth for each user. 
Similarly, we assume that the GBS also adopts the equal bandwidth allocation scheme, i.e., each non-cell-edge MT $k\in\mathcal{K}_G$ is allocated with an effective bandwidth of $b_G W$, with $b_G\triangleq(1-\rho)/K_G$. 
On the other hand, we also assume a similar equal bandwidth allocation scheme in the spectrum reuse case, despite that the total bandwidth is now used by both the UAV and GBS concurrently. 

In the following two sections, we will present the two spectrum sharing models in more details as well as their respective design optimization problems and solutions to maximize the system common throughput.

\section{Orthogonal Spectrum Sharing}\label{SectionFormulation}

In this section, we study the orthogonal spectrum sharing scheme. First, we derive the achievable throughput of the UAV-MT and GBS-MT communications, respectively. Denote the common (minimum) throughput of all MTs as $\bar\nu$ in bits per second per Hz (bps/Hz), which is normalized with respect to the total system bandwidth $W$.
Then, we formulate the problem to maximize $\bar\nu$ by jointly optimizing the UAV trajectory radius $r_U$, user partitioning radius threshold $r_I$, and bandwidth allocation portion $\rho$.

\subsection{UAV-MT Communication}\label{OrthogonalUAV}

\subsubsection{Average throughput}
For each MT $k$, we define the \textit{association time} $\tau_k$ as the total time duration in which MT $k$ is associated with the UAV for communications within each UAV flying period $T$.
The average throughput of cell-edge MT $k\in\mathcal{K}_U$ over each period $T$ is determined by $\tau_k$ and its instantaneous communication rate with the UAV during this association time interval.

Assume that the UAV allocates transmit power $p_k(t)$ to communicate with MT $k\in\mathcal{K}_U(t)$ at time $t$ during its association time. Then the instantaneous achievable rate $R_k(t)$ of MT $k\in\mathcal{K}_U(t)$ in bps/Hz is given by 
\begin{equation}\label{RateUAVinst1}
R_k(t)=b_U(t)\log_2\bigg( 1+ \frac{G_U h_{k}(t) p_k(t)}{b_U(t) \sigma^2}\bigg)=b_U(t)\log_2\bigg( 1+ \frac{\eta_0 G_U p_k(t)}{b_U(t) \big(d_k^2(t)+H_U^2\big)}\bigg),
\end{equation}
where the receiver noise is assumed to be additive white Gaussian with power spectrum density $N_0$ in Watts/Hz; $\sigma^2\triangleq N_0W$ is the total noise power over the whole bandwidth of $W$ Hz; and $\eta_0\triangleq\beta_0/\sigma^2$.
It can be seen that $R_k(t)$ is determined by the allocated transmit power $p_k(t)$, the UAV-MT horizontal link distance $d_{k}(t)$, and the normalized per-user bandwidth $b_U(t)$ which in turn depends on the number of MTs $|\mathcal{K}_U(t)|$ associated with the UAV at time $t$.

With \eqref{RateUAVinst1}, the average throughput of cell-edge MT $k\in\mathcal{K}_U$ within a UAV flying period $T$ is given by 
\begin{equation}\label{RateUAVavg1}
\bar R_k=\frac{1}{T}\int_{t=t_{s,k}}^{t_{e,k}} R_k(t)\diff t,
\end{equation}
where $t_{s,k}$ and $t_{e,k}$ are the starting and ending time instants for the interval when MT $k$ is associated with the UAV, respectively, and $\tau_k=t_{e,k}-t_{s,k}$.
Next, we discuss the design of transmit power $p_k(t), t_{s,k}\leq t\leq t_{e,k}$, the UAV-MT association $\mathcal{K}_U(t), 0\leq t\leq T$, and the distance $d_{k}(t), t_{s,k}\leq t\leq t_{e,k}$, respectively.

\subsubsection{Power allocation}
Let $P_U$ denote the maximum transmit power of the UAV.
For simplicity, we assume that at each time instant $t$, the UAV allocates equal transmit power to its associated MTs $k\in\mathcal{K}_U(t)$, i.e., $p_k(t)=P_U/|\mathcal{K}_U(t)|, \forall k\in\mathcal{K}_U(t)$.
From \eqref{RateUAVinst1} and using the fact that $b_U(t)=\rho/|\mathcal{K}_U(t)|$, the instantaneous achievable rate $R_k(t)$ becomes
\begin{equation}\label{RateUAVinstEP}
R_k(t)=b_U(t)\log_2\bigg( 1+ \frac{\eta_0 G_U P_U/|\mathcal{K}_U(t)|}{b_U(t) \big(d_k^2(t)+H_U^2\big)}\bigg)=\frac{\rho}{|\mathcal{K}_U(t)|}\log_2\bigg( 1+ \frac{\eta_0 G_U P_U}{\rho \big(d_k^2(t)+H_U^2\big)}\bigg),
\end{equation}
which depends on $\rho$, $G_U$, $d_{k}(t)$ and $|\mathcal{K}_U(t)|$. 
The association $\mathcal{K}_U(t), 0\leq t\leq T$ determines the average throughput $\bar R_k$ in \eqref{RateUAVavg1} in two ways, namely, the normalized per-user bandwidth $b_U(t)=\rho /|\mathcal{K}_U(t)|$ at each time $t$, and the association time period $t_{s,k}\leq t\leq t_{e,k}$ assigned for each MT $k$.

\subsubsection{UAV-MT association}\label{SectionAssociation}

\begin{figure}
\centering
   \includegraphics[width=0.4\linewidth]{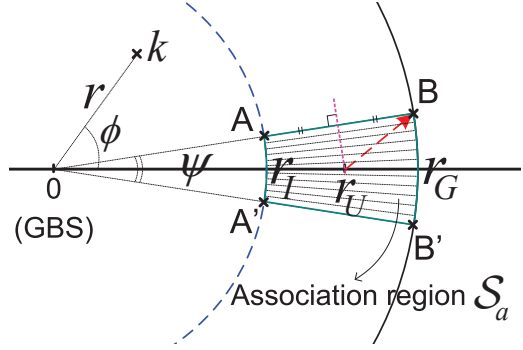}
\caption{Proposed UAV-MT association pattern.\vspace{-2ex}}\label{UAVGBSassociation}
\end{figure}

For the analytical tractability, we design a simple yet practical UAV-MT association rule as follows.
At each time $t$, assume that the horizontal position of the UAV is at $(r_U, 0)$ in the polar coordinate system $(r,\phi)$.
The MTs $k\in\mathcal{K}_U$ in the ring region with $r_I\leq r\leq r_G$ are to be served by the UAV via cyclical multiple access. Accordingly, we choose a \textit{ring segment} region (denoted as $\mathcal{S}_a$) with central angle $\psi$, which is also symmetric about the horizontal axis, as shown by the shadowed region in Fig. \ref{UAVGBSassociation}.
Within the region $\mathcal{S}_a$, any arc centered at the origin (GBS location) with radius $r_I\leq r\leq r_G$ has the same central angle $\psi$. 
In particular, denote the arcs with radius $r_I$ and $r_G$ by AA' and BB', respectively.

We propose the UAV-MT association rule by which the MTs within the ring segment region $\mathcal{S}_a$ are associated with the UAV for communications at time $t$, which thus determines the set $\mathcal{K}_U(t)$.
This association rule simplifies our subsequent analysis in two aspects. Firstly, all cell-edge MTs $k\in\mathcal{K}_U$ have equal association time with the UAV, i.e.,
\begin{equation}\label{AssociationTime}
\tau_k=\frac{\psi T}{2\pi},\forall k\in \mathcal{K}_U.
\end{equation}
Secondly, the average number of MTs associated with the UAV at any time $t$ is a linearly increasing function of $\psi$, i.e., 
\begin{equation}\label{AssociationNumber}
K_a\triangleq\lambda S_a=\lambda (r_G^2-r_I^2)\psi/2,
\end{equation}
where $S_a\triangleq (r_G^2-r_I^2)\psi/2$ is the area of $\mathcal{S}_a$. 

Note that with the proposed association rule, 
each MT $k\in\mathcal{K}_U$ incurs an \textit{access delay} \cite{CyclicalLyu} given by $D_k\triangleq T-\tau_k$, which is the time duration within each UAV flying period $T$ when MT $k$ is not associated with the UAV for communications. 
Therefore, the proposed scheme is most suitable for the cell-edge MTs with high throughput demand but less stringent delay requirement. For those cell-edge MTs with stringent delay requirement, it can still be served by the GBS in the conventional way.
On the other hand, 
the cell-edge MTs are exclusively served by the UAV in a cyclical time-division manner, while the non-cell-edge MTs are exclusively served by the GBS. In other words, there is no need for handover of any MT between the GBS and UAV.

\subsubsection{Lower bound of average throughput}

Based on the above association rule, the association time $\tau_k$ in \eqref{AssociationTime} is identical for all MTs $k\in\mathcal{K}_U$. Therefore, the average throughput $\bar R_k$ in \eqref{RateUAVavg1} is determined by
the instantaneous rate $R_k(t), t_{s,k}\leq t\leq t_{e,k}$, which depends on $\rho$, $G_U$, $d_{k}(t)$ and $b_U(t)$.
In the following, we derive a lower bound for the average throughput $\bar R_k$ in \eqref{RateUAVavg1}, based on the upper bound of the UAV-MT horizontal distance $d_{k}(t)$ and the lower bound of normalized per-user bandwidth $b_U(t)$.

First, $d_{k}(t)$ is a non-linear function of $t$ and it is different for MTs located at different $r$. 
Denote $d_{\textrm{max}}$ as the upper bound of the horizontal distance from the UAV to any point in the ring segment region $\mathcal{S}_a$.
Since $\mathcal{S}_a$ should lie within the coverage area of the UAV, we have $r_c\geq d_{\textrm{max}}$, i.e., $H_U\tan\Phi_U\geq d_{\textrm{max}}$, which yields
\begin{equation}\label{rcConstraint}
\Phi_U\geq \arctan(d_{\textrm{max}}/H_U).
\end{equation}
Since the UAV's antenna gain of the main lobe $G_U$ in \eqref{UAVantenna} is a decreasing function of $\Phi_U$, $\Phi_U$ should be chosen to be the minimum possible value as in \eqref{rcConstraint} in order to maximize $G_U$ and hence the throughput.
Therefore, the UAV antenna gain $G_U$ towards the coverage area is given by
\begin{equation}\label{GUdmax}
G_U(d_{\textrm{max}})=\frac{G_0}{(\arctan\frac{d_{\textrm{max}}}{H_U} )^2},
\end{equation}
which is a decreasing function of $d_{\textrm{max}}$.

It can be verified that $d_{\textrm{max}}$ always occurs at one of the two intersection points A and B as shown in Fig. \ref{UAVGBSassociation}. Denote $d_{A}$ and $d_{B}$ as the horizontal distances from the UAV to points A and B, respectively. Then we have
\begin{equation}\label{dmax}
d_{\textrm{max}}= \max(d_{A},d_{B}),
\end{equation}
where $d_{A}$ and $d_{B}$ can be obtained by using the cosine law as follows
\begin{equation}\label{da}
d_{A}=\sqrt{r_U^2 +r_I^2-2r_U r_I\cos\frac{\psi}{2}},
\end{equation}
\begin{equation}\label{db}
d_{B}=\sqrt{r_U^2 +r_G^2-2r_U r_G\cos\frac{\psi}{2}}.
\end{equation}
It can be verified that $d_{\textrm{max}}$ is an increasing function of $\psi$ for any given $r_I$ and $r_U$.

Second, let $K_{a,\textrm{max}}\triangleq \max\limits_{0\leq t\leq T} |\mathcal{K}_U(t)|$ denote the maximum number of MTs associated with the UAV over the period $T$, and denote $\mu\triangleq \frac{K_{a,\textrm{max}}}{K_a}\geq 1$.
Note that $\mu$ depends on the spatial variations of the user locations. 
Then at any time $t$, $b_U(t)$ is lower-bounded by
\begin{equation}\label{buLower}
b_U(t)\geq  \frac{\rho}{K_{a,\textrm{max}}}=\frac{2\rho}{\mu \lambda (r_G^2-r_I^2)\psi}\triangleq b_{\textrm{min}},
\end{equation}
where the lower bound $b_{\textrm{min}}$ is inversely proportional to $\psi$.

Then the instantaneous rate $R_k(t)$ in \eqref{RateUAVinstEP} for any MT $k\in\mathcal{K}_U(t)$ at any time $t$ is lower-bounded by
\begin{equation}\label{RateUAVinstEPlower}
R_k(t)\geq b_{\textrm{min}}\log_2\bigg( 1+ \frac{\eta_0 P_U G_U(d_{\textrm{max}})}{\rho (d_{\textrm{max}}^2+H_U^2)}\bigg)\triangleq  R_U,
\end{equation}
where the lower bound $R_U$ is a decreasing function of $\psi$, since a larger central angle $\psi$ leads to larger $d_{\textrm{max}}$ and smaller $b_{\textrm{min}}$.

Based on \eqref{RateUAVinstEPlower}, we then assume that the UAV communicates with each MT $k\in\mathcal{K}_U(t)$ at any time $t$ using a constant rate equal to $R_U$, which is achievable for all MTs in $\mathcal{K}_U(t)$. Then the average throughput in \eqref{RateUAVavg1} for MT $k\in\mathcal{K}_U$ over each time period $T$ is given by
\begin{equation}\label{RateUAVavgLB}
\bar R_k=\frac{\tau_k}{T} R_U=\frac{\psi}{2\pi} R_U,
\end{equation}
which is equal for every cell-edge MT $k\in\mathcal{K}_U$. 
Therefore, by substituting $R_U$ from \eqref{RateUAVinstEPlower} and $b_{\textrm{min}}$ from \eqref{buLower} into \eqref{RateUAVavgLB}, the common throughput $\bar R_U$ for the cell-edge MTs served by the UAV can be expressed as
\begin{align}
\bar R_U(\rho, r_I, d_{\textrm{max}})&\triangleq\frac{\psi}{2\pi} R_U =\frac{\psi}{2\pi}b_{\textrm{min}}\log_2\bigg( 1+ \frac{\eta_0 P_U G_U(d_{\textrm{max}})}{\rho (d_{\textrm{max}}^2+H_U^2)}\bigg)\notag\\ 
             &\stackrel{(a)}{=}\frac{\rho}{\mu \lambda \pi(r_G^2-r_I^2)}\log_2\bigg( 1+ \frac{\eta_0 P_U G_U(d_{\textrm{max}})}{\rho (d_{\textrm{max}}^2+H_U^2)}\bigg),\label{RateUAVavgLBu}
\end{align}
which is a function of $\rho, r_I$ and $d_{\textrm{max}}$. Note that the equality $(a)$ follows since the proportional effect of $\psi$ on the association time $\tau_k$ in \eqref{AssociationTime} cancels out its inversely proportional effect on the per-user bandwidth $b_{\textrm{min}}$ in \eqref{buLower}, under our proposed association rule.

Since $\bar R_U$ decreases with $d_{\textrm{max}}$ which in turn increases with $\psi$, it is desirable to choose $\psi$ as small as possible to increase $\bar R_U$ in \eqref{RateUAVavgLBu}. However, $\psi$ cannot be arbitrarily small in practice, since there might be no MTs associated with the UAV at some time $t$, i.e., $|\mathcal{K}_U(t)|=0$. In the rest of this paper, we assume that the value of $\psi$ is given, and hence the corresponding $d_{\textrm{max}}$ can be obtained based on \eqref{dmax}--\eqref{db}, which is a function of $r_I$ and $r_U$. Therefore, \eqref{RateUAVavgLBu} becomes
\begin{equation}\label{RateUAVavgLBuNew}
\bar R_U(\rho, r_I, r_U)=\frac{\rho}{\mu \lambda \pi(r_G^2-r_I^2)}\log_2\bigg( 1+ \frac{\eta_0 P_U G_U(d_{\textrm{max}})}{\rho (d_{\textrm{max}}^2+H_U^2)}\bigg).
\end{equation}

Finally, we define the \textit{spatial throughput} as the aggregated throughput per unit area in bps/Hz/m$^2$, i.e., $\theta\triangleq\frac{\sum R_k}{S}$, where $S$ is the area of interest. The spatial throughput of the UAV-served area is thus given by $\theta_U\triangleq\lambda\bar R_U(\rho, r_I, r_U)$, i.e.,
\begin{equation}\label{SpatialThroughputUAV}
\theta_U=\frac{\rho}{\mu \pi(r_G^2-r_I^2)}\log_2\bigg( 1+ \frac{\eta_0 P_U G_U(d_{\textrm{max}})}{\rho (d_{\textrm{max}}^2+H_U^2)}\bigg).
\end{equation}

\subsection{GBS-MT Communication}

On the other hand, the MTs inside the inner disk of radius $r_I$ are associated with the GBS for communications, which form the non-cell-edge MT set $\mathcal{K}_G$.
Recall that the GBS-MT channel gain $g_k$ consists of the average channel gain $\bar g_k$ which depends on the GBS-MT horizontal distance $r$ with $r\leq r_I$, and an additional random term $\zeta_k\sim \textrm{Exp}(1)$ accounting for small-scale fading of the channel. 
We assume that the GBS knows the average channel gain $\bar g_k$ for each MT $k$ and the distribution of $\zeta_k$.

\subsubsection{Power allocation}\label{OrthogonalGBSpower}
Assume that the GBS transmits with equal power $p_G(r)$ for MTs at the same distance $r$ from the GBS, with $r\leq r_I$. 
We consider that the GBS adopts the ``slow"  channel inversion power control \cite{goldsmith2005wireless} based on the average channel gain $\bar g_k$ (instead of the instantaneous channel gain which requires the estimation of the instantaneous channels and hence is more costly for practical implementation), i.e., the transmit power $p_G(r)$ is allocated such that all MTs $k\in\mathcal{K}_G$ have the equal \textit{average} signal-to-noise ratio (SNR) at the receiver, denoted by $\bar\gamma$.
Thus, $p_G(r)$ can be expressed as
\begin{equation}\label{ConstraintGBSpowerSNR}
p_G(r)=\frac{\bar\gamma b_G \sigma^2}{\bar g_k G_G}=\frac{\bar\gamma b_G (H_G^2+r^2)^{\frac{n}{2}}}{\kappa_0}, \forall r, 0\leq r\leq r_I,
\end{equation}
where $\kappa_0\triangleq\alpha_0 G_G/\sigma^2$, and the allocated power $p_G(r)$ is inversely proportional to the average channel gain $\bar g_k$.

Let $P_G$ denote the maximum transmit power of the GBS. 
Then the total transmit power to all MTs associated with the GBS needs to satisfy the following constraint:
\begin{equation}\label{ConstraintGBSpower}
\lambda\int_{\phi=0}^{2\pi}\int_{r=0}^{r_I} p_G(r)r\diff r\diff \phi= P_G.
\end{equation}
The average SNR can be obtained from \eqref{ConstraintGBSpowerSNR} and \eqref{ConstraintGBSpower} as
\begin{equation}\label{Gamma}
\bar\gamma=\frac{\kappa_0 P_G}{2\pi\lambda b_G L(r_I)}=\frac{\kappa_0 P_G r_I^2}{2 (1-\rho) L(r_I)},
\end{equation}
where $b_G=\frac{1-\rho}{\lambda\pi r_I^2}$ and 
\begin{equation}\label{L}
L(r_I)\triangleq\int_{r=0}^{r_I} (H_G^2+r^2)^{\frac{n}{2}} r\diff r=\frac{(H_G^2+r_I^2)^{\frac{2+n}{2}}-H_G^{2+n}}{2+n}.
\end{equation}
The instantaneous achievable rate for MT $k\in\mathcal{K}_G$ in bps/Hz is then given by
\begin{equation}\label{RateGBSinst}
R_k=b_G\log_2( 1+ \bar\gamma \zeta_{k}).
\end{equation}

\subsubsection{Outage probability}
Due to the small-scale fading of the GBS-MT channel, an outage event occurs when the GBS-MT link cannot support the desired common throughput $\bar\nu$.
According to \eqref{RateGBSinst}, the outage probability for MT $k\in\mathcal{K}_G$ is given by
\begin{align}
\textrm{P}_{\textrm{out},k}&=\textrm{Pr}\{b_G\log_2( 1+ \bar\gamma \zeta_{k})<\bar\nu\}=\textrm{Pr}\{\zeta_{k}<(2^{\bar\nu/b_G}-1)/\bar\gamma\}\notag\\
&=1-\exp \big(-(2^{\bar\nu/b_G}-1)/\bar\gamma\big)\triangleq \textrm{P}_{\textrm{out}}(\rho,r_I,\bar \nu),\label{OutageProb1}
\end{align}
which is equal for all MTs $k\in\mathcal{K}_G$ due to the common average SNR $\bar\gamma$ with the adopted channel inversion power control.
For convenience,
define a function $f(\rho,r_I,\bar \nu)$ as follows:
\begin{equation}\label{OutageFunction}
f(\rho,r_I,\bar \nu)\triangleq \frac{2^{\bar\nu/b_G}-1}{\bar\gamma}=\frac{2\big(2^{\frac{\pi r_I^2\cdot\lambda\bar\nu }{1-\rho}}-1\big) (1-\rho) L(r_I)}{\kappa_0 P_G r_I^2}.
\end{equation}
Then we have 
\begin{equation}\label{OutageProbFunction}
\textrm{P}_{\textrm{out}}(\rho,r_I,\bar \nu)=1-\exp \big(-f(\rho,r_I,\bar \nu)\big).
\end{equation}
It can be verified from \eqref{OutageFunction} that $f(\rho,r_I,\bar \nu)$ and hence $\textrm{P}_{\textrm{out}}(\rho,r_I,\bar \nu)$ are both increasing functions of $\rho$, $r_I$ and $\bar \nu$.

Define $\theta_G\triangleq\lambda\bar\nu$ as the spatial throughput of the GBS-served area.
Suppose that the allowed maximum outage probability is $\bar{\textrm{P}}_{\textrm{out}}$ for all GBS-MT links.
Note that in the special case without the UAV, i.e., $\rho=0$ and $r_I=r_G$, by letting $\textrm{P}_{\textrm{out}}(\rho=0,r_I=r_G,\bar \nu)=\bar{\textrm{P}}_{\textrm{out}}$ in \eqref{OutageProbFunction}, we can then obtain the common throughput $\bar\nu_G^{\textrm{opt}}$ and the corresponding spatial throughput for all MTs in this case.

\subsection{Problem Formulation}

In this subsection, we formulate the optimization problem to maximize the common throughput $\bar\nu$ of all MTs subject to the maximum outage probability constraint of GBS-MT links, by jointly optimizing the bandwidth allocation portion $\rho$, the user partitioning distance threshold $r_I$, and the UAV trajectory radius $r_U$.
The problem can be formulated as
\begin{align}
\mathrm{(P1)}: \underset{
\begin{subarray}{c}
  \rho, r_I, r_U, \bar\nu\\
  \end{subarray}
}{\max}& \quad\bar\nu \notag\\
             \text{s.t.}\quad&\textrm{P}_{\textrm{out}}(\rho,r_I,\bar \nu)\leq \bar{\textrm{P}}_{\textrm{out}},\label{ConstraintOutage}\\ 
             &\bar R_U(\rho,r_I,  r_U)\geq \bar\nu,\label{ConstraintRU}\\ 
             &r_I\leq r_U\leq r_G,\label{ConstraintrU}\\ 
             &0\leq r_I\leq r_G,\label{ConstraintrI}\\ 
             &0\leq \rho\leq 1.\label{Constraintrho}
\end{align}
We denote the optimal solution to (P1) as $(\rho^{\textrm{opt}}, r_I^{\textrm{opt}}, r_U^{\textrm{opt}})$ and the corresponding optimal common throughput as $\bar\nu^{\textrm{opt}}$.

\subsection{Proposed Solution}\label{SectionSolution}
Solving problem (P1) is non-trivial due to the non-convex constraints \eqref{ConstraintOutage} and \eqref{ConstraintRU}. By exploiting its special structure, (P1) is \textit{optimally} solved as follows.

First, (P1) can be equivalently reduced to a series of sub-problems (P2) given below, each for a given target value $\bar\nu$. Furthermore, $\bar\nu$ can be updated via bisection search method.
Specifically, to check whether a certain $\bar\nu$ is achievable, we can solve problem (P2) which minimizes the outage probability of GBS-MT links subject to the constraints \eqref{ConstraintRU}--\eqref{Constraintrho}, i.e.,
\begin{align}
\mathrm{(P2)}:\underset{
\begin{subarray}{c}
  \rho, r_I, r_U\\
  \end{subarray}
}{\min}&\quad \textrm{P}_{\textrm{out}}(\rho, r_I,\bar\nu) \notag\\
             \text{s.t.}\quad&\quad\eqref{ConstraintRU}\textrm{ -- }  \eqref{Constraintrho}. \notag
\end{align}
If the optimal value of (P2) is no larger than $\bar{\textrm{P}}_{\textrm{out}}$, then \eqref{ConstraintOutage} is satisfied, and the optimal solution to (P2) and the corresponding $\bar\nu$ is a feasible solution to (P1).
On the other hand, if the optimal value of (P2) is larger than $\bar{\textrm{P}}_{\textrm{out}}$, then the corresponding $\bar\nu$ value is not achievable.
Accordingly, bisection search can be applied to find the maximum common throughput $\bar\nu^{\textrm{opt}}$ iteratively. We thus focus on solving (P2) in the following.

Second, (P2) is still difficult to be directly solved, due to the non-convex objective function and the non-convex constraint \eqref{ConstraintRU}. 
Fortunately,
since the GBS-MT communication is independent of $r_U$, with given fixed $\rho$ and $r_I$, we can first optimize $r_U$ to maximize the achievable UAV-MT common throughput $R_U(\rho, r_I, r_U)$ while satisfying the constraint \eqref{ConstraintrU}, i.e.,
\[
\mathrm{(P3)}:
\underset{
\begin{subarray}{c}
  r_I\leq r_U\leq r_G\\
  \end{subarray}
}{\max}\quad \bar R_U(\rho, r_I, r_U).
\]
Denote the optimal value of (P3) as $\bar R_U^{\textrm{max}}(\rho,r_I)$. Problem (P3) can be optimally solved based on geometry as detailed in Section \ref{SectionOptrU} below.

Third, after $\bar R_U^{\textrm{max}}(\rho,r_I)$ is obtained from (P3), problem (P2) can be equivalently reduced to the following problem.
\begin{align}
\mathrm{(P4)}:
\underset{
\begin{subarray}{c}
  \rho,r_I\\
  \end{subarray}
}{\min}&\quad f(\rho, r_I,\bar\nu) \notag\\
             \text{s.t.}&\quad\eqref{ConstraintrI}\textrm{ and }  \eqref{Constraintrho}, \notag\\
       &\quad\bar\nu-\bar R_U^{\textrm{max}}(\rho,r_I)\leq 0, \label{ConstraintUAVrate2}     
\end{align}
where the objective function $\textrm{P}_{\textrm{out}}(\rho, r_I,\bar\nu)$ of (P2) is replaced by $f(\rho, r_I,\bar\nu)$ based on monotonicity in \eqref{OutageProbFunction}, and the constraint \eqref{ConstraintRU} is replaced by \eqref{ConstraintUAVrate2}.

Finally, by exploiting the monotonicity of the objective function and constraint function over $\rho$ and $r_I$, (P4) can be optimally solved by bi-section searching for $\rho$ in the range $0<\rho<1$ for given $r_I$ in the inner loop, while performing a one-dimensional search for $r_I$ in the range $0\leq r_I\leq r_G$ in the outer loop. The details are provided in Section \ref{SectionRhorI} below.

\subsubsection{Optimizing $r_U$}\label{SectionOptrU}
To solve (P3) for given $\rho$ and $r_I$, we need to maximize $\bar R_U(\rho, r_I, r_U)$ in \eqref{RateUAVavgLBuNew} by optimizing $r_U$, which is equivalent to minimizing $d_{\textrm{max}}=\max(d_A, d_B)$ given by \eqref{dmax}, \eqref{da} and \eqref{db}. 
For $r_I\leq r_U\leq r_G$ and a given small value $\psi\leq \psi_0$ ($\psi_0$ will be derived later), the minimum $d_{\textrm{max}}$ can be found by letting $d_A=d_B$ in \eqref{da} and \eqref{db}, which yields 
\begin{equation}\label{rU2}
r_{U}^*=\frac{r_G+r_I}{2\cos (\psi/2)},
\end{equation}
and
\begin{equation}\label{dmax22}
d_{\textrm{max}}^*(r_I)=\sqrt{\frac{(r_G+r_I)^2}{2(\cos\psi+1)}-r_I r_G},
\end{equation}
where $d_{\textrm{max}}^*(r_I)$ is a decreasing function of $r_I$.
Note that the coordinate $(r_U^*,0)$ corresponds to the intersection point of the horizontal axis and the perpendicular bisector of the line segment AB, as shown in Fig.~\ref{UAVGBSassociation}. 
By geometry, it can be verified that when $r_U=r_U^*$, the minimum value of $d_{\textrm{max}}$ is achieved as that given by \eqref{dmax22}.
This conclusion is valid when the coordinate $(r_U^*,0)$ does not go beyond the mid-point $(r_G\cos\frac{\psi}{2},0)$ of the line segment BB', since otherwise the minimum value of $d_{\textrm{max}}$ simply equals half the length of the line segment BB', i.e., $r_G\sin\frac{\psi}{2}$. Therefore, from 
$\frac{r_G+r_I}{2\cos (\psi/2)}\leq r_G\cos\frac{\psi}{2}$, we obtain the threshold $\psi_0$ as follows.
\begin{equation}\label{Contraintpsi111}
\psi_0\triangleq \arccos\frac{r_I}{r_G}<\frac{\pi}{2}.
\end{equation}

By substituting $d_{\textrm{max}}=d_{\textrm{max}}^*(r_I)$ in \eqref{RateUAVavgLBuNew}, we obtain the optimal value of (P3) which is given by
\begin{equation}\label{RUmax}
\bar R_U^{\textrm{max}}(\rho,r_I)=
\frac{\rho}{\mu \lambda \pi(r_G^2-r_I^2)}\log_2\bigg( 1+ \frac{\eta_0 P_U G_U\big(d_{\textrm{max}}^*(r_I)\big)}{\rho \big(\big(d_{\textrm{max}}^*(r_I)\big)^2+H_U^2\big)}\bigg).
\end{equation}
It can be verified that 
$\bar R_U^{\textrm{max}}(\rho,r_I)$ is an increasing function of both $\rho$ and $r_I$.


\subsubsection{Optimizing $\rho$ and $r_I$}\label{SectionRhorI}
Next, we investigate the performance trade-off between GBS-MT and UAV-MT communications by optimizing $\rho$ and $r_I$ in (P4).
In general, a larger $\rho$ means that more bandwidth is allocated to the UAV, thus improving the max-min throughput of UAV-MT communications but at the cost of degrading that of GBS-MT communications.
On the other hand, a larger $r_I$ means that more MTs are to be served by the GBS, which also degrades the max-min throughput of GBS-MT communications while improving that of UAV-MT communications.

Specifically, given $\bar \nu$ in (P4), the objective function $f(\rho, r_I,\bar\nu)$ (defined in \eqref{OutageFunction}) is a non-convex function of either $\rho$ or $r_I$. Moreover, the constraint in \eqref{ConstraintUAVrate2} is a non-convex constraint since $\bar\nu-\bar R_U^{\textrm{max}}(\rho,r_I)$ (with $\bar R_U^{\textrm{max}}(\rho,r_I)$ given in \eqref{RUmax}) is non-convex with respect to $r_I$. Therefore, (P4) is a non-convex optimization problem and thus cannot be directly solved with the standard convex optimization techniques.

Fortunately, we can exploit the monotonicity of $\bar R_U^{\textrm{max}}(\rho,r_I)$ and $f(\rho, r_I,\bar\nu)$ with $\rho$ and $r_I$ to devise an efficient algorithm to solve (P4) optimally as follows.
It is observed that given $\bar\nu$ and $r_I$, the objective function $f(\rho, r_I,\bar\nu)$ increases with $\rho$ while the constraint function $\bar\nu-\bar R_U^{\textrm{max}}(\rho,r_I)$ decreases with $\rho$. Therefore, in order to minimize $f(\rho, r_I,\bar\nu)$, we should choose the minimum value of $\rho$ that satisfies the constraint \eqref{ConstraintUAVrate2}. Since $\bar\nu-\bar R_U^{\textrm{max}}(\rho,r_I)$ decreases with $\rho$, a bisection search for $\rho$ in the range of $0<\rho<1$ can be performed to check the feasibility of the constraint \eqref{ConstraintUAVrate2}, and to find the minimum $\rho$ if feasible. Then, we can perform a one-dimensional search for the optimal $r_I$ in the range of $0\leq r_I\leq r_G$ to further minimize the objective function $f(\rho, r_I,\bar\nu)$ in (P4).

\section{Spectrum Reuse}\label{SectionReuse}

In this section, we extend our analysis to the spectrum reuse scheme where the common spectrum pool of total bandwidth $W$ Hz is shared by both the GBS and UAV, which is expected to further improve the spectrum efficiency as long as the mutual interference is well controlled between the UAV-MT and GBS-MT communications.
To this end, we propose to leverage the joint use of directional/sectorized antennas at the UAV/GBS to eliminate the mutual interference and thus maximize the throughput performance.
Since there is no need to design $\rho$ in the spectrum reuse case, we focus on   the joint optimization of the UAV trajectory radius $r_U$ and the user partitioning distance threshold $r_I$ to maximize the minimum throughput $\bar\nu$ of all MTs.

\subsection{GBS-MT Communication}

\subsubsection{Directional transmission}
\begin{figure}
\centering
   \includegraphics[width=0.55\linewidth]{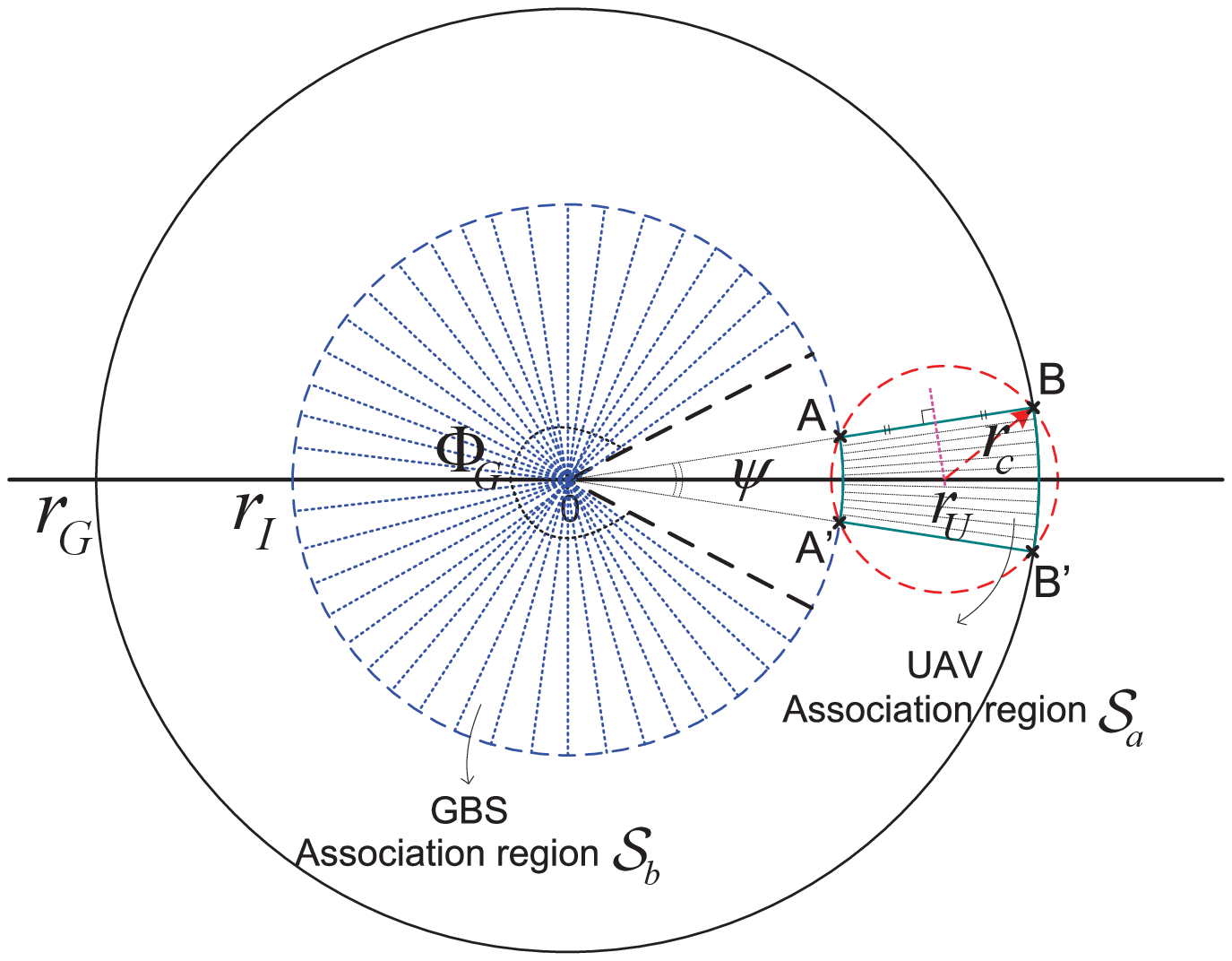}
\caption{Proposed spectrum reuse model with interference-free concurrent cyclical multiple access for both UAV-MT and GBS-MT communications.\vspace{-2ex}}\label{UAVGBSsector}
\end{figure}

As shown in Fig. \ref{UAVGBSsector}, we assume that the GBS dynamically adjusts its transmission direction towards the shadowed sector region $\mathcal{S}_b$ with central angle $\Phi_G$, which is non-overlapping with the central angle $\psi$ of the UAV association region $\mathcal{S}_a$ at each time, and thus causes no interference to the UAV-MT communications. 
Assume that the GBS antenna gain in the $\Phi_G$ direction remains as $G_G$ for fair comparison with the orthogonal sharing case.
We further assume that the non-cell-edge MTs in $\mathcal{S}_b$ are associated with the GBS for communications at time $t$, denoted by the set $\mathcal{K}_G(t)\in \mathcal{K}_G$.
Then on average there are $|\mathcal{K}_G(t)|=\lambda r_I^2\Phi_G/2$ MTs in $\mathcal{K}_G(t)$. Assume that the GBS also adopts the simple equal bandwidth allocation scheme, i.e., each MT in $\mathcal{K}_G(t)$ is allocated with an effective normalized bandwidth $b_G(t)=1/|\mathcal{K}_G(t)|=2/(\lambda r_I^2\Phi_G)$.

Thanks to the directional antenna at the UAV, there is practically negligible interference from the UAV to the GBS-MT communications as well.
As the UAV flies cyclically, the GBS adapts its transmission direction accordingly, which can be implemented by adaptive beamforming techniques or approximately by on-off control of the sectorized antennas in practice. 
As a result, the GBS-MT communications also become cyclical multiple access with the same period $T$ as the UAV-MT communications, where each MT $k\in\mathcal{K}_G$ has an access delay $D_k=(1-\frac{\Phi_G}{2\pi})T$.

\subsubsection{Power allocation}

At time $t$, assume that the GBS adopts the ``slow"  channel inversion power control similar to Section \ref{OrthogonalGBSpower}, despite that the associated MTs become $\mathcal{K}_G(t)$ instead.
Assume that the GBS transmits with the same power $p_G(r)$ for MTs $k\in\mathcal{K}_G(t)$ at the same distance $r$ from the GBS. 
The transmit power $p_G(r)$ is allocated such that all MTs $k\in\mathcal{K}_G(t)$ have the equal average SNR at the receiver, denoted by $\bar\gamma(t)$.
Thus, $p_G(r)$ can be expressed as
\begin{equation}\label{ConstraintGBSpowerSNRR}
p_G(r)=\frac{\bar\gamma(t) b_G(t) \sigma^2}{\bar g_k G_G}=\frac{\bar\gamma(t) b_G(t) (H_G^2+r^2)^{\frac{n}{2}}}{\kappa_0}. 
\end{equation}

Let $P_G$ denote the maximum transmit power of the GBS. 
Then the total transmit power to all MTs in $\mathcal{K}_G(t)$ needs to satisfy the following constraint:
\begin{equation}\label{ConstraintGBSpowerR}
\lambda\int_{\phi=0}^{\Phi_G}\int_{r=0}^{r_I} p_G(r)r\diff r\diff \phi= P_G.
\end{equation}
The average SNR can be obtained from \eqref{ConstraintGBSpowerSNRR} and \eqref{ConstraintGBSpowerR} as
\begin{equation}\label{GammaR}
\bar\gamma(t)=\frac{\kappa_0 P_G}{\Phi_G\lambda b_G(t) L(r_I)}=\frac{\kappa_0 P_G  r_I^2}{2 L(r_I)},
\end{equation}
where $L(r_I)$ is given by \eqref{L}.
The instantaneous achievable rate for MT $k\in\mathcal{K}_G(t)$ in bps/Hz is then given by
\begin{equation}\label{RateGBSinstR}
R_k(t)=b_G(t)\log_2\big( 1+ \bar\gamma(t) \zeta_{k}\big).
\end{equation}

\subsubsection{Outage probability}
Due to the small-scale fading of the GBS-MT channel, an outage event occurs when the GBS-MT link cannot support the desired instantaneous rate $\bar\nu_G\triangleq \frac{2\pi}{\Phi_G}\bar\nu$, where $\bar\nu$ is the desired average throughput in a period $T$.
According to \eqref{RateGBSinstR}, the outage probability for MT $k\in\mathcal{K}_G(t)$ is given by
\begin{align}
\textrm{P}_{\textrm{out},k}(t)&=\textrm{Pr}\big\{b_G(t)\log_2\big( 1+ \bar\gamma(t) \zeta_{k}\big)<\bar\nu_G\big\}\notag\\
&=\textrm{Pr}\bigg\{\frac{2}{\lambda r_I^2\Phi_G} \log_2\big( 1+ \bar\gamma(t) \zeta_{k}\big)<\frac{2\pi}{\Phi_G}\bar\nu\bigg\}\notag\\
&=\textrm{Pr}\big\{\zeta_{k}<(2^{\pi r_I^2\cdot\lambda\bar\nu}-1)/\bar\gamma(t)\big\}\notag\\
&=1-\exp \big(-(2^{\pi r_I^2\cdot\lambda\bar\nu}-1)/\bar\gamma(t)\big)\notag\\
&=1-\exp \bigg(\frac{-2(2^{\pi r_I^2\cdot\lambda\bar\nu}-1) L(r_I)}{\kappa_0 P_G  r_I^2}\bigg)\triangleq \textrm{P}_{\textrm{out}}'(r_I,\bar \nu),\label{OutageProb}
\end{align}
which is identical for all MTs $k\in\mathcal{K}_G(t)$.
It can be verified from \eqref{OutageProb} that $\textrm{P}_{\textrm{out}}'(r_I,\bar \nu)$ is an increasing function of $r_I$ and $\bar \nu$.
Note that $\textrm{P}_{\textrm{out}}'(r_I,\bar \nu)$ is equal to $\textrm{P}_{\textrm{out}}(\rho,r_I,\bar \nu)$ in \eqref{OutageProbFunction} with $\rho=0$, i.e., when the whole bandwidth is used by the GBS.
Since $\textrm{P}_{\textrm{out}}(\rho,r_I,\bar \nu)$ is an increasing function of $\rho$, the outage probability decreases to its minimum value when $\rho=0$. Therefore, the spectrum reuse scheme has a lower outage probability than that of the orthogonal sharing scheme under the same $r_I$ and $\bar\nu$, which implies a higher throughput achievable by the spectrum reuse scheme under the same outage requirement.
Finally, note that the central angle $\Phi_G$ does not affect $\textrm{P}_{\textrm{out}}'(r_I,\bar \nu)$, which can thus be chosen in practice to be as large as possible to reduce the user access delay, provided that the leakage interference to the UAV-MT communications is kept sufficiently low.

\subsection{UAV-MT Communication}
Since the interference from the GBS is eliminated, the UAV-MT communication is similar to that in Section \ref{OrthogonalUAV}, but the whole bandwidth is now used by the UAV. Therefore, the common throughput $\bar R_U'$ for the cell-edge MTs served by the UAV follows from \eqref{RateUAVavgLBuNew} with $\rho=1$, i.e.,
\begin{equation}\label{RateUAVavgLBuNewR}
\bar R_U'(r_I, r_U)=\frac{1}{\mu \lambda \pi(r_G^2-r_I^2)}\log_2\bigg( 1+ \frac{\eta_0 P_U G_U(d_{\textrm{max}})}{d_{\textrm{max}}^2+H_U^2}\bigg).
\end{equation}
which is a function of $r_I$ and $r_U$.

\subsection{Problem Formulation}

In this subsection, we formulate the optimization problem to maximize the common throughput $\bar\nu$ of all MTs subject to the maximum outage probability constraint of GBS-MT links, by jointly optimizing the user partitioning distance threshold $r_I$, and the UAV trajectory radius $r_U$.
The problem can be formulated as
\begin{align}
\mathrm{(P5)}: \underset{
\begin{subarray}{c}
  r_I, r_U, \bar\nu\\
  \end{subarray}
}{\max}& \quad\bar\nu \notag\\
             \text{s.t.}\quad&\textrm{P}_{\textrm{out}}'(r_I,\bar \nu)\leq \bar{\textrm{P}}_{\textrm{out}},\label{ConstraintOutageR}\\ 
             &\bar R_U'(r_I,  r_U)\geq \bar\nu,\label{ConstraintRUR}\\ 
             &r_I\leq r_U\leq r_G,\label{ConstraintrUR}\\ 
             &0\leq r_I\leq r_G.\label{ConstraintrIR}
\end{align}
We denote the optimal solution to (P5) as $(r_I^{\textrm{opt'}},r_U^{\textrm{opt'}})$ and the corresponding optimal common throughput as $\bar\nu^{\textrm{opt'}}$.
Note that (P5) is similar to (P1), except that the bandwidth partition between the UAV and GBS is no more needed.

\subsection{Proposed Solution}
Problem (P5) can be solved using similar methods as in Section \ref{SectionSolution}.
First, for any given $r_I$, the UAV trajectory radius $r_U$ can be optimized to achieve the maximum UAV-MT throughput, denoted as $\bar R_U'^{\textrm{max}}(r_I)$, which, by following Section \ref{SectionOptrU}, is given by
\begin{equation}\label{RUmaxR}
\bar R_U'^{\textrm{max}}(r_I)=\frac{1}{\mu \lambda \pi(r_G^2-r_I^2)}\log_2\bigg( 1+ \frac{\eta_0 P_U G_U\big(d_{\textrm{max}}^*(r_I)\big)}{\big(d_{\textrm{max}}^*(r_I)\big)^2+H_U^2}\bigg),
\end{equation}
where the optimal $r_U$ follows from \eqref{rU2} and $d_{\textrm{max}}^*(r_I)$ is given by \eqref{dmax22}.
It can be verified that 
$\bar R_U'^{\textrm{max}}(r_I)$ is an increasing function of $r_I$.

Second, for any given $r_I$, the maximum GBS-MT throughput, denoted as $\bar R_G'^{\textrm{max}}(r_I)$, can be found as $\bar\nu$ when the constraint \eqref{ConstraintOutageR} holds with equality. It can be verified that 
$\bar R_G'^{\textrm{max}}(r_I)$ is a decreasing function of $r_I$.
Finally, we can perform a bisection search to find the optimal $r_I$, which achieves the max-min throughput $\bar\nu^{\textrm{opt'}}=\max\limits_{r_I}\min\{\bar R_U'^{\textrm{max}}(r_I), \bar R_G'^{\textrm{max}}(r_I)\}$.

Note that the proposed spectrum reuse scheme requires adaptive directional transmissions at the GBS and cyclical multiple access for the GBS-MT communications, which thus requires additional complexity for implementation. However, thanks to the interference avoidance, the GBS and UAV can both access the common spectrum pool for concurrent communications, which thus further improves the system throughput, as will be shown in the next section.

\subsection{Further Discussions}\label{SectionDiscuss}
\subsubsection{Relaxation of fixed UAV altitude}
For the schemes proposed above, the optimization results provide useful guidelines to practically design the UAV trajectory radius $r_U$, bandwidth allocation portion $\rho$, and user partitioning distance threshold $r_I$, which jointly determine the radius $r_c$ of the UAV coverage area so that the scheduled MTs in $\mathcal{K}_U (t)$ communicating with the UAV are guaranteed to lie within the UAV coverage area at any time $t$.

We have assumed that the UAV has a given altitude $H_U$ for simplicity. In the case where the UAV flies at different altitudes along its optimized trajectory, the UAV antenna beamwidth $\Phi_U$ can be adjusted accordingly to achieve the same coverage area of any fixed radius $r_c=H_U\tan\Phi_U$, and thus there is no fundamental change of our results with variable altitude. 
Moreover, the coverage radius $r_c$ is only a theoretical upper bound to guarantee that all currently scheduled MTs communicating with the UAV lie within the coverage area. Therefore, a certain level of altitude/beamwidth control error in practice can be tolerated for our proposed design.

\subsubsection{Requirement of User Location Information}
In this paper, we mainly target for outdoor scenarios with temporary hot spot, where our proposed optimization schemes only require the statistics of the user distribution instead of the exact location of each ground user. The obtained results provide a theoretical guideline to design the UAV trajectory, bandwidth allocation, and user partitioning in practice. Assume that the UAV follows the optimized trajectory to serve the MTs within its ground coverage area at each time instant, where the scheduled MTs typically have high received power from the UAV. As a result, the served MTs over time can be determined by using the reference signal received power (RSRP). Although accurate location information of the ground MTs can be a plus, it is not a must for our proposed schemes.

More specifically, we assume that the MTs are uniformly and randomly distributed by following a homogeneous Poisson point process (HPPP) with a certain density $\lambda$, where $\lambda$ is constant in the considered cell. 
Therefore, a hotspot cell occurs when the user density $\lambda$ is large. 
Under this model, a circular UAV trajectory with a constant speed along the cell edge effectively shortens the communication distance from the UAV to its associated cell-edge users, thus improving the system overall throughput.
On the other hand, in scenarios where there exist ``hotspots in hotspot" and the specific locations of such non-uniformly distributed users are known, the UAV trajectory and flying speed can be optimized to further improve the throughput performance. For example, the UAV can fly closer to or hover above ``hotspots in hotspot" so as to shorten the communication distance and/or maintain a longer communication duration for the users therein to improve the throughput.

\subsubsection{Extension to Multiple Cells}
In this paper, as a preliminary study of our proposed new network architecture, we focus on the single-cell setup to investigate the fundamental design issues such as the UAV trajectory, spectrum sharing and multiple access, while the significant performance gain shown for the single-cell case will be the motivation for us to investigate UAV-aided cellular offloading for the more general multi-cell case in future work. 
Here we briefly discuss the possible extensions to the multi-cell setup.

Firstly, when multiple UAVs are available for a single cell, the additional UAVs can be arranged along the designed trajectory with equal separation from each other, which helps reduce the access delay and also improve the throughput. Secondly, the results developed in the current paper can be directly applied when a single UAV is available for each cell in the multi-cell scenario. There are various possible ways to mitigate the interference between a UAV flying along the cell edge and its neighboring cells. For example, in the current paper we have considered the use of a directional antenna at the UAV, which already limits the interference to/from neighboring cells. 
Another issue is the collision avoidance between UAVs serving adjacent cells. Fortunately, in the current paper the optimized UAV trajectory lies inside the cell boundary, which theoretically avoids the collision among UAVs in different cells.
Thirdly, when a single UAV is responsible for serving multiple cells, the UAV can be scheduled to serve these cells sequentially, for each of them following circular or other optimized trajectories, which is worth studying in future work. 
Finally, when a macro cell is overloaded and yet there are relatively few micro BSs (small cells) nearby, the UAV can still be employed to help offload data traffic from the macro BS. In such a case, the specific design of the UAV offloading scheme, including the UAV trajectory, user partitioning and spectrum sharing, needs to take into account the locations, user partitioning and spectrum sharing of the existing macro BS and micro BSs, which deserves further investigation.

\subsubsection{Energy Constraint for UAVs}
In practice, UAVs usually have limited endurance due to on-board energy constraint. One potential solution for it is to employ multiple UAVs that take turns to provide service and recharge/swap battery on the ground. Thanks to emerging techniques such as automated battery swap and recharge \cite{UAVbatterySwap}, a single UAV can accomplish long-endurance missions by automatically swapping its depleted battery at a ground charging station with a fully charged battery. Moreover, in the case with fixed-wing UAVs \cite{FixedWingUAV}, their flight endurance is typically much longer than that of rotary-wing UAVs, which can be several hours and thus are suitable for our considered application.
In Section \ref{SectionEE}, a quantitative example is provided for the energy efficiency of UAV-aided communication in our considered system.

On the other hand, the proposed UAV-assisted offloading scheme is mainly targeted for the scenarios of temporary hotspot where the existing ground infrastructure is incapable of serving the suddenly-surged traffic demand, and it is practically costly or takes too long to install new ground infrastructure to meet such high demand. In these cases, UAVs can be more swiftly deployed in the target area to provide high throughput for the ground MTs temporally. Therefore, with the above methods for prolonging the UAV flight endurance, the proposed UAV-assisted cellular offloading scheme offers a viable new approach to resolve the hot-spot issue in the forthcoming 5G and beyond.

\section{Numerical Results}\label{SectionSimulation}

In this section, numerical results are provided to validate our analysis and evaluate the performance of our proposed schemes, which consist of two parts. In the first part, we evaluate the performance of our proposed schemes with optimized and fixed design parameters, respectively, and also compare with the benchmark scheme with GBS only.
In the second part, the proposed scheme is further compared with the conventional cell-edge throughput enhancement scheme which deploys one or more micro/small cells at the edge of the macro cell.

\subsection{Performance Evalution of the Proposed Schemes}\label{SimulationPart1}

For the orthogonal sharing scheme,
we obtain the optimal solution $(\rho^{\textrm{opt}}, r_I^{\textrm{opt}}, r_U^{\textrm{opt}})$ to (P1) with the maximum common throughput $\bar\nu^{\textrm{opt}}$ and corresponding maximum spatial throughput $\theta^{\textrm{opt}}=\lambda\bar\nu^{\textrm{opt}}$. We compare the spatial throughput with those of two benchmark schemes. The first benchmark considers fixed design variables with $\rho=0.5$, $r_I/r_G=0.5$ and $r_U$ following \eqref{rU2}, where the spatial throughput is taken to be the minimum throughput of the GBS- and UAV-served areas, i.e., $\theta^{\textrm{fixed}}\triangleq\min(\theta_G,\theta_U)$. The second benchmark considers the GBS-only case without the use of UAV.
On the other hand, for the spectrum reuse scheme, we obtain the optimal solution $(r_I^{\textrm{opt'}},r_U^{\textrm{opt'}})$ to (P5) with the maximum common throughput $\bar\nu^{\textrm{opt'}}$ and corresponding maximum spatial throughput $\theta^{\textrm{opt'}}=\lambda\bar\nu^{\textrm{opt'}}$. We also compare with the benchmark scheme with fixed design variable $r_I/r_G=0.5$ and $r_U$ following \eqref{rU2}.

For each of these schemes, the obtained analytical results are verified by averaging over 100 independent realizations of the user locations.
Each realization is drawn from a homogeneous Poisson point process (HPPP) with the given user density $\lambda$. 
In each realization, the GBS channel inversion power control is simulated based on specific user locations, while the parameter $\mu$ for UAV-MT association can be obtained as the average value over the 100 realizations for our analytical results. 
We then obtain the average spatial throughputs $\bar\theta_G$ and $\bar\theta_U$ for the GBS- and UAV-served areas over the 100 realizations, respectively. 
The following parameters are used: $f_c=2$ GHz, $W=10$ MHz, $N_0=-174$ dBm/Hz, $H_U=100$ m, $H_G=20$ m, $r_G=1000$ m, $G_G=16$ dBi, $n=3$, $\psi=\frac{\pi}{6}$, $\Phi_G=\frac{4\pi}{3}$ and $\bar{\textrm{P}}_{\textrm{out}}=0.01$.

\begin{figure}
\centering
\includegraphics[width=0.8\linewidth,  trim=65 0 65 0,clip]{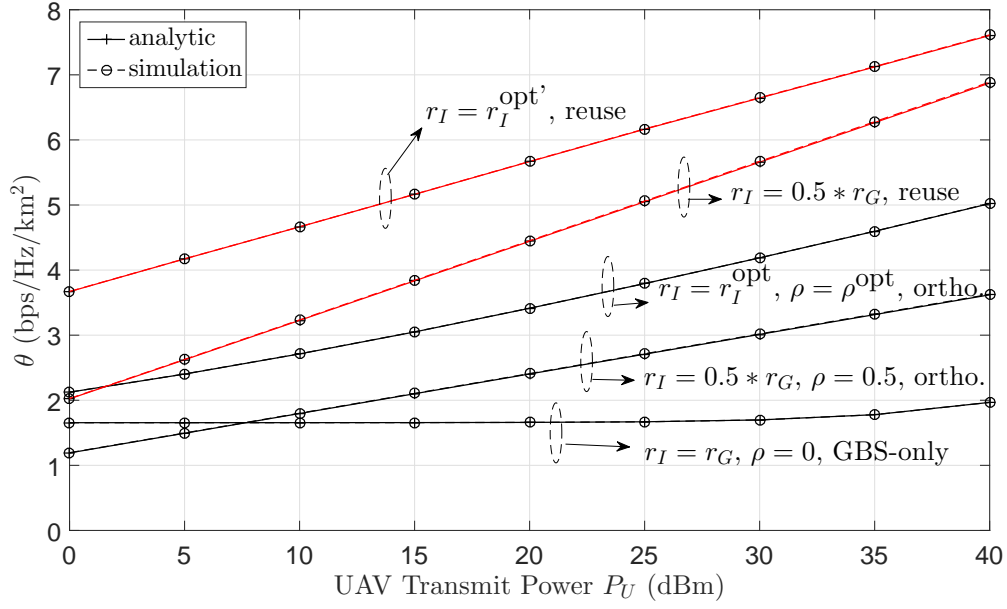} 
\caption{Spatial throughput $\theta$ under different UAV transmit power $P_U$.\vspace{-2ex}
} \label{PUcurve}
\end{figure}

\begin{figure}
\centering
\includegraphics[width=0.7\linewidth,  trim=65 0 65 0,clip]{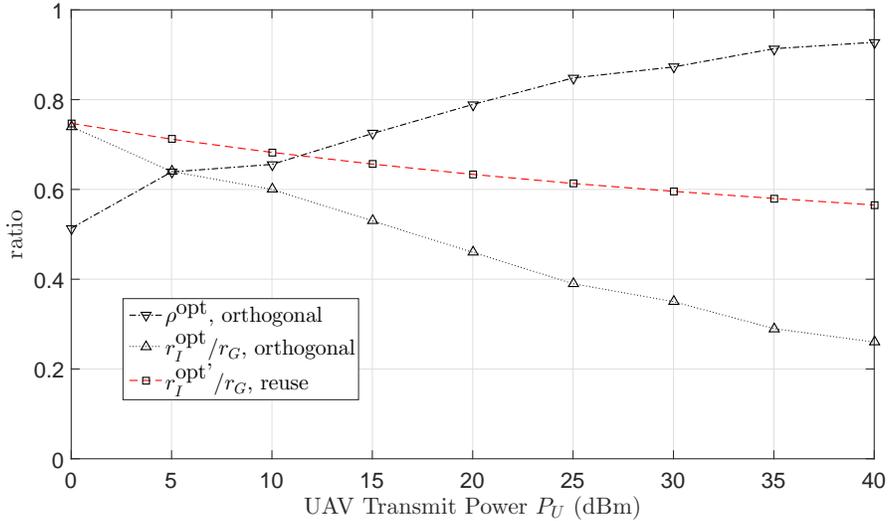} 
\caption{Optimal solutions of orthogonal sharing versus spectrum reuse under different UAV transmit power $P_U$.\vspace{-2ex}
} \label{PUcurverI}
\end{figure}

In the first set of simulations, we choose $\lambda=1000$ MTs/km$^2$ and $P_G=40$ dBm, 
and simulate the above schemes with different UAV transmit power $P_U$, where the UAV's available transmit power $P_U$ is added to the GBS transmit power $P_G$ in the GBS-only benchmark case for fair comparison.
The throughput results are plotted in Fig. \ref{PUcurve}, and the optimal solutions to (P1) and (P5) are plotted in Fig. \ref{PUcurverI}, respectively.
First, it can be observed
from Fig. \ref{PUcurve} that the analytical results match well with the simulation results in all cases.
Second, for the orthogonal sharing case, our proposed scheme even with fixed (unoptimized) $\rho$ and $r_I$ improves the spatial throughput over the GBS-only case when $P_U\geq 10$ dBm. On the other hand, our proposed scheme with optimized $\rho$ and $r_I$ further improves over the case with fixed $\rho$ and $r_I$, and achieves the maximum spatial throughput which is significantly higher than that of the GBS-only case for all $P_U$ values.
Moreover, as $P_U$ increases, it can be seen from Fig. \ref{PUcurverI} that $\rho^{\textrm{opt}}$ increases and $r_I^{\textrm{opt}}/r_G$ decreases for the orthogonal sharing scheme, which suggests that more bandwidth should be allocated to the UAV to serve more MTs when the UAV is able to transmit at a higher power.
In contrast, for the spectrum reuse case, it can be seen from Fig. \ref{PUcurve} that our proposed scheme with optimized or fixed $r_I$ further improves the spatial throughput significantly as compared to the corresponding orthogonal sharing case.
It is also noted from Fig. \ref{PUcurverI} that the optimal solution $r_I^{\textrm{opt'}}$ in the spectrum reuse scheme decreases as $P_U$ increases, which suggests that more users should be served by the UAV when the UAV is able to transmit at higher power.
Moreover, $r_I^{\textrm{opt'}}$ is larger than $r_I^{\textrm{opt}}$ in the orthogonal sharing scheme as shown in Fig. \ref{PUcurverI}, since the GBS in the spectrum reuse case is able to use more bandwidth and thus should serve more non-cell-edge users to achieve the maximum common throughput.
In summary, our proposed joint optimization solution is essential to achieve the maximum throughput of the proposed UAV-assisted hybrid network. 

\begin{figure}
\centering
\includegraphics[width=0.8\linewidth,  trim=40 0 65 0,clip]{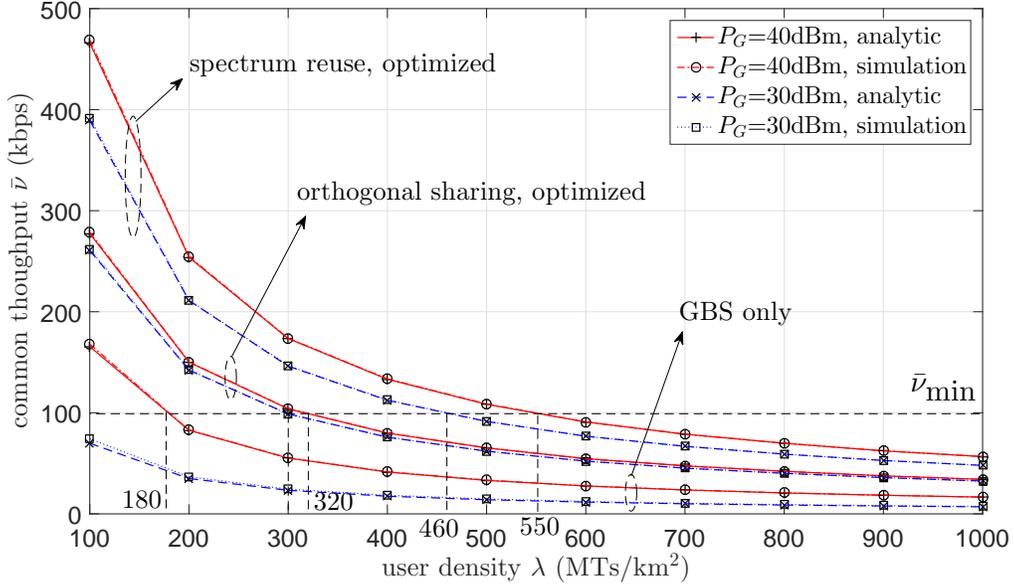} 
\caption{Common throughput $\bar\nu$ under different user density $\lambda$.\vspace{-2ex}
} \label{curveLambda}
\end{figure}

To illustrate the offloading performance more explicitly, in the second set of simulations, we compare the maximum user density $\lambda_{\textrm{max}}$ that can be supported by various schemes under the constraint that the common throughput per MT $\bar\nu$ should be no less than a minimum required value $\bar\nu_{\textrm{min}}$.
To this end, we consider the orthogonal sharing and spectrum reuse schemes with their respective optimized designs, and compare the obtained common throughput $\bar\nu$ with that of the GBS-only case under different user density $\lambda$.
We choose $P_U=20$ dBm and $P_G=30$ or 40 dBm, and the results are plotted in Fig. \ref{curveLambda}.
First, it can be observed
from Fig. \ref{curveLambda} that the analytical results match well with the simulation results in all cases.
Second, the common throughput $\bar\nu$ decreases as the user density $\lambda$ increases in all cases, since the limited resource is shared by more users.
Third, suppose that the minimum desired throughput is $\bar\nu_{\textrm{min}}=100$ kbps, then we can find the maximum user density $\lambda_{\textrm{max}}$ supported by each scheme. In the GBS-only case, we have $\lambda_{\textrm{max}}<100$ MTs/km$^2$ for the case with $P_G=30$ dBm, and the density further increases to $\lambda_{\textrm{max}}=180$ MTs/km$^2$ with a larger transmit power $P_G=40$ dBm.
In the optimized orthogonal sharing scheme, $\lambda_{\textrm{max}}=300$ and 320 MTs/km$^2$ for the cases with $P_G=30$ dBm and $P_G=40$ dBm, respectively, which significantly outperforms the conventional system with GBS only.
With the optimized spectrum reuse scheme, the maximum supported user density further increases to $\lambda_{\textrm{max}}=460$ and 550 MTs/km$^2$ for the cases with $P_G=30$ dBm and $P_G=40$ dBm, respectively, which offers more performance gains over the optimized orthogonal sharing scheme.
In summary, our proposed orthogonal sharing and spectrum reuse schemes with optimal designs can support higher user density than the GBS-only case, which shows the great potential of our proposed UAV-aided cellular offloading to address the cellular hotspot issue.

\subsection{Illustrative Example of UAV Energy Efficiency}\label{SectionEE}
The energy efficiency is another important aspect of UAV-aided communication. In this subsection, we provide a simple example to illustrate how to evaluate UAV energy efficiency in the proposed design.

An initial attempt for quantifying the energy efficiency of UAV-enabled communication is given in \cite{ZengEnergyEffcient}, where the energy efficiency is defined as the amount of transmitted information bits per unit energy (Joule) consumed by the UAV, which accounts for the UAV's dominant propulsion energy consumption. For fixed-wing UAVs with level flight under normal operations, a generic energy consumption model is proposed in \cite{ZengEnergyEffcient}, which takes into account the UAV's instantaneous velocity and acceleration. In particular, for the UAV with a constant flying speed $V$ following a circular trajectory with radius $r_U$, the propulsion power is modeled in \cite{ZengEnergyEffcient} as
\begin{equation}\label{Pcir}
\bar{P}_\textrm{cir}(V,r_U)=\bigg(c_1+\frac{c_2}{g^2 r_U^2}\bigg)V^3+\frac{c_2}{V},
\end{equation}
where $g=9.8$ m/s$^2$ is the gravitational acceleration, $c_1$ is a modeling coefficient to account for the parasitic power for overcoming the parasitic drag due to the aircraft's skin friction and form drag, and $c_2$ is a modeling coefficient to account for the induced power for overcoming the lift-induced drag. As can be seen from \eqref{Pcir}, the UAV power consumption of a circular trajectory decreases with $r_U$, and for given $r_U$, there is an optimum speed $V^*$ at which the power consumption is minimized.

In our setup, the total UAV transmitted information bits within a UAV flying period $T$ can be estimated as $B=T W\pi(r_G^2-r_I^2 )\theta_U$, where $W$ is the system bandwidth and $\theta_U$ is the obtained spatial throughput over the UAV-served area. Consider an example setup of the orthogonal sharing scheme with $\rho=0.5$, $r_I=0.5r_G$, $\lambda=1000$ MTs/km$^2$, $\psi=\pi/6$, $P_U=1$ W and $c_1=9.26\times 10^{-4}, c_2=2250$ as given in \cite{ZengEnergyEffcient}, while the rest of the parameters are given in Section \ref{SimulationPart1}. The optimized UAV trajectory radius follows from \eqref{rU2} and is given by $r_U^*=(r_G+r_I)/\big(2 \cos(\psi/2)\big)=776$ m. The obtained spatial throughput is given by $\theta_U\approx 3.0$ bps/Hz/km$^2$. The optimum speed at which the power consumption is minimized is given by $V^*=29.7$ m/s \cite{ZengEnergyEffcient}, while the corresponding UAV propulsion power follows from \eqref{Pcir} and is given by $\bar{P}_\textrm{cir}(V^*,r_U^*)=101.03$ W. The overall energy efficiency of UAV communication is thus given by
\begin{equation}
\textrm{EE}=\frac{B}{T\big(P_U+\bar{P}_\textrm{cir}(V^*,r_U^*)\big)}=\frac{W\pi(r_G^2-r_I^2 )\theta_U}{P_U+\bar{P}_\textrm{cir}(V^*,r_U^*)}=693 \textrm{ kbits/Joule}.
\end{equation}

\subsection{Comparison with Micro-Cell Offload Scheme}

\begin{figure}
        \centering
        \begin{subfigure}[b]{0.4\linewidth}
                \includegraphics[width=1\linewidth,  trim=210 20 210 30,clip]{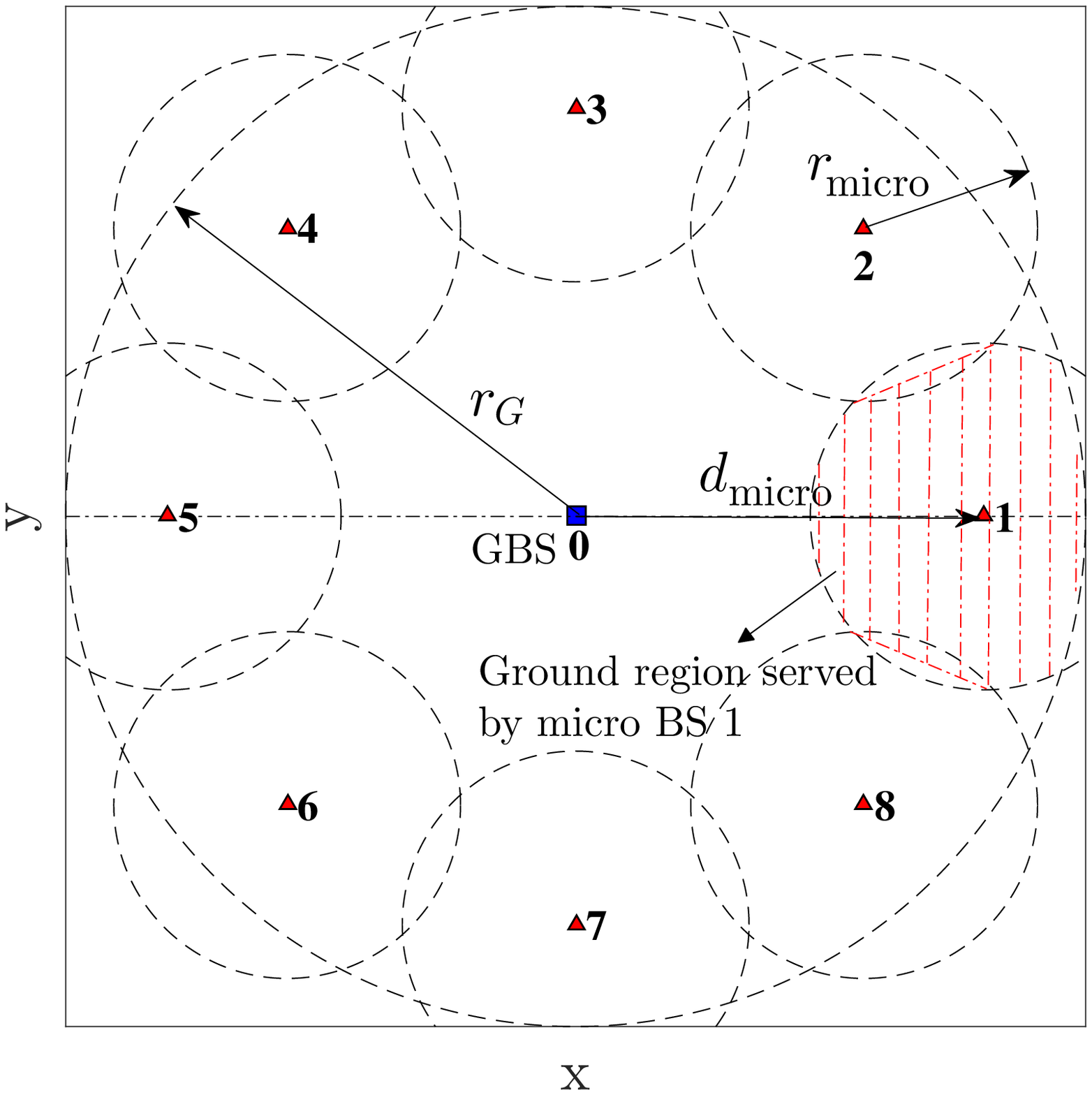}
                \caption{$M=8$}
                \label{FigureMicro8}
        \end{subfigure}%
        \begin{subfigure}[b]{0.4\linewidth}
                \includegraphics[width=1\linewidth,  trim=220 20 220 30,clip]{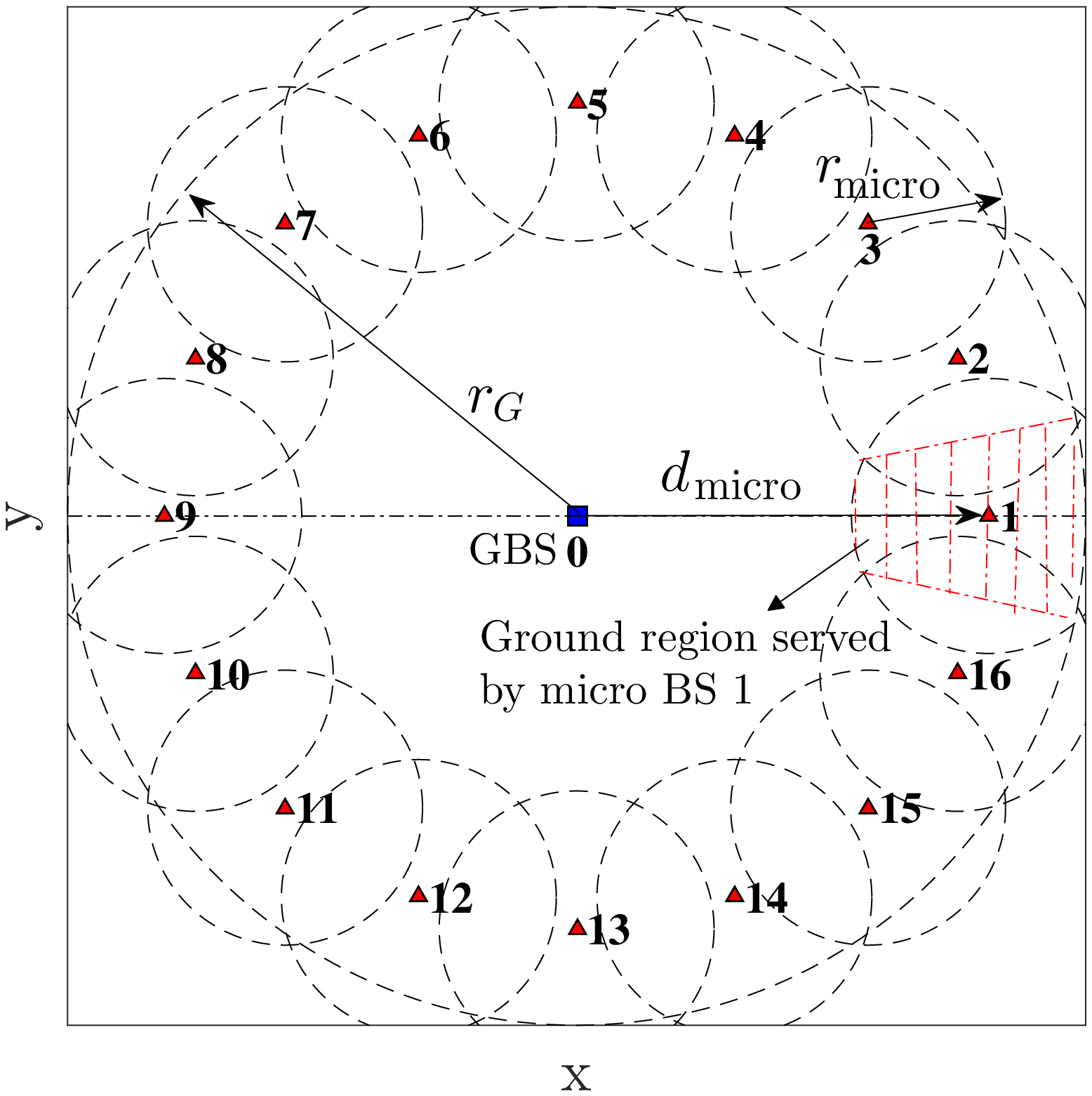}
                \caption{$M=16$}
                \label{FigureMicro16}
        \end{subfigure}%
        \caption{Benchmark scheme with $M$ micro-cells at the cell edge.}\label{FigureMicro}
\end{figure}

In this subsection, the proposed scheme is further compared with the conventional cell-edge throughput enhancement scheme which deploys micro/small cells at the edge of the macro cell.
Specifically, we consider the benchmark scheme where $M$ micro-cell BSs are uniformly placed at a distance $d_\textrm{micro}$ from the GBS at the origin, which help to offload data traffic from MTs in the macro cell with radius $r_G$. Examples for $M=8$ and $M=16$ are shown in Fig. \ref{FigureMicro} (a) and (b), respectively. Denote $r_\textrm{micro}$ as the radius of the disk coverage region of each micro BS, which helps to serve the MTs within its coverage region. In the case where the coverage regions of two adjacent micro BSs overlap, an MT in the overlapping region is served by its nearest micro BS. For example, the ground region served by micro BS 1 is represented by the shadowed area in Fig. \ref{FigureMicro}. The remaining MTs in the macro cell which are not covered by any micro BS are associated with the GBS for communication.

Assume that each micro BS is equipped with an omnidirectional antenna at height $H_\textrm{micro}$ with antenna gain $G_\textrm{micro}$. The channel between the micro BSs and the MTs is modeled similarly as to that of the GBS-MT channels. For simplicity, we investigate the case with orthogonal spectrum sharing between the GBS and the micro BSs. Assume that a portion $\rho_\textrm{micro}$ ($0<\rho_\textrm{micro}<1$) of the total bandwidth $W$ is allocated to the micro BSs, where each micro BS is allocated with an equal portion of $\rho_\textrm{micro}/M$. Further assume that both the GBS and the micro BSs adopt the equal bandwidth sharing for their associated MTs, respectively. Assume that the total transmit power of the micro BSs is $P_\textrm{micro}$, where each micro BS has an equal transmit power of $P_\textrm{micro}/M$. Further assume that both the GBS and the micro BSs adopt the “slow” channel inversion power control (similar to that in Section \ref{OrthogonalGBSpower}) based on the average channel gain of their associated MTs, respectively.

\begin{figure}
\centering
\includegraphics[width=0.8\linewidth,  trim=0 0 0 0,clip]{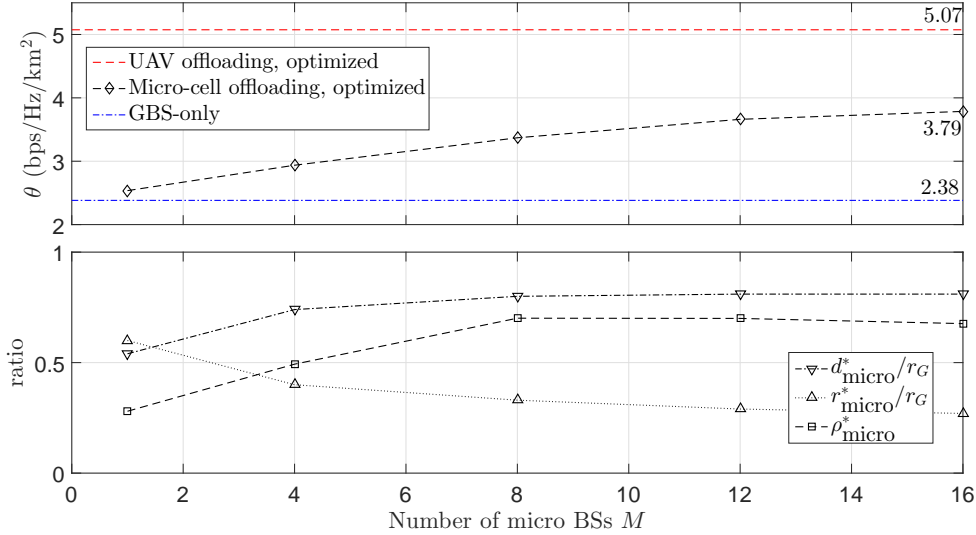} 
\caption{Spatial throughput $\theta$ in the micro-cell offloading scheme with optimized parameters $d_\textrm{micro}^*$, $r_\textrm{micro}^*$ and $\rho_\textrm{micro}^*$.\vspace{-2ex}
} \label{MicroCurve}
\end{figure}

Under the above setup, in the simulations we can independently generate $N=20$ realizations of the MT locations which follow the HPPP distribution with given user density $\lambda$. For each realization, we obtain numerical results for the average throughput $\nu_G$ and $\nu_\textrm{micro}$ of the MTs served by the GBS and the micro BSs, respectively, for given $d_\textrm{micro}$, $r_\textrm{micro}$ and $\rho_\textrm{micro}$. Then for given $d_\textrm{micro}$ and $r_\textrm{micro}$, we exhaustively search for the optimal $\rho_\textrm{micro}$ to maximize the minimum throughput $\nu=\min\{\nu_G,\nu_\textrm{micro}\}$ for a given setup, and then obtain the average throughput $\bar\nu$ by averaging over the $N$ realizations. Then we exhaustively search for the optimal $d_\textrm{micro}$ and $r_\textrm{micro}$ to maximize $\bar\nu$, and obtain the corresponding spatial throughput $\theta$. The maximum spatial throughput and the corresponding optimal solutions $d_\textrm{micro}^*$, $r_\textrm{micro}^*$ and $\rho_\textrm{micro}^*$ for $M=1,4,8,12$ and 16 are plotted in Fig. \ref{MicroCurve}, respectively. The parameter values from Section \ref{SimulationPart1} are used here except the following: $\lambda=1000$ MTs/km$^2$, $H_\textrm{micro}=10$ m, $G_\textrm{micro}=8$ dBi, $P_\textrm{micro}=40$ dBm and $P_G=46$ dBm.

For comparison, in Fig. \ref{MicroCurve} we also show the maximum spatial throughput obtained by our optimized UAV offloading scheme where the UAV has transmit power $P_U=P_\textrm{micro}$, as well as that obtained by the GBS-only scheme where $P_\textrm{micro}$ is added to the transmit power $P_G$ of the GBS for fair comparison. It can be seen that the spatial throughput obtained by the micro-cell offloading scheme gradually increases as the number of micro cells increases, where the optimized micro-cell placement and layout tend to be pushed closer to the cell edge and thus able to achieve better offloading performance compared to the benchmark scheme with the GBS only. Nevertheless, the proposed UAV offloading scheme with only one single UAV/mobile BS still significantly outperforms the micro-cell offloading scheme in terms of throughput for all values of $M$. The above performance gain is mainly due to the fact that the UAV in general offers better communication links to its served ground MTs due to the LoS channels and its mobility.

\section{Conclusions}\label{SectionConclusion}

This paper proposes a new hybrid network architecture for cellular systems by leveraging the use of UAVs for data offloading.
We first investigate the orthogonal spectrum sharing scheme between the UAV and GBS, and solve the problem to maximize the common throughput of all MTs in the cell by jointly optimizing the spectrum allocation, user partitioning, and UAV trajectory design.
We then extend our study to the spectrum reuse scheme where the common spectrum pool is shared by both the GBS and UAV while effectively suppressing their mutual interference via adaptive directional transmissions, which further improves the spatial throughput.
Numerical results show that the proposed hybrid network design significantly improves the throughput as compared to the conventional system with the GBS only. 
Moreover, our optimized UAV offloading scheme with only one single UAV is shown to be able to significantly outperform the conventional cell-edge throughput enhancement scheme with multiple micro/small cells in terms of throughput, besides saving the infrastructure cost.
We hope that this work would lead to a new practical solution to address the hotspot issue in future 5G and beyond-5G wireless systems. There are still many important issues unsolved in the proposed new hybrid wireless network, e.g., how to extend this work to the scenarios with multiple UAVs and/or multiple cells is challenging and worth investigating in future work. 


\bibliography{IEEEabrv,BibDIRP}

\newpage

\end{document}